\documentclass{article}

\usepackage{arxiv}

\usepackage[utf8]{inputenc} 
\usepackage[T1]{fontenc}    
\usepackage{hyperref}       
\usepackage{url}            
\usepackage{booktabs}       
\usepackage{amsfonts}       
\usepackage{nicefrac}       
\usepackage{microtype}      
\usepackage{lipsum}
\usepackage{soul,color}
\usepackage{graphicx}
 
\title{Big Data driven Product Design: A Survey}

\author{
  Huafeng Quan\thanks{Correspondence.} \\
  College of Big Data and Statistics\\
  Guizhou Universify of Finance and Economics\\
  Guiyang, 550050, China \\
  \texttt{huafengquan@mail.gufe.edu.cn} \\
   \And
Shaobo Li \\
  State Key Laboratory of Public Big Data\\
  Guizhou University\\
  Guiyang, 550050, China \\
  \texttt{lishaobo@gzu.edu.cn} \\
    \And
Changchang Zeng \\
  Chengdu Institute of Computer Application\\
  Chinese Academy of Sciences\\
  Chengdu, 610041, China\\
  \texttt{zengchangchang16@mails.ucas.ac.cn} \\
     \And
Hongjing Wei \\
  School of Mechanical Engineering\\
  Guizhou Institute of Technology\\
  Guiyang, 550050, China\\
  \texttt{hongjingwei@126.com} \\
  \And
Jianjun Hu \\
  Department of Computer Science and Engineering\\
  University of South Carolina\\
  Columbia, SC, 29201, USA \\
  \texttt{jianjunh@cse.sc.edu} \\
}

\begin{document}
\maketitle

\begin{abstract}
With the improvement of living standards, user requirements of modern products are becoming increasingly more diversified and personalized. Traditional product design methods can no longer satisfy the market needs due to their strong subjectivity, small survey scope, poor real-time data, and lack of visual display, which calls for the development of big data driven product design methodology. Big data in the product lifecycle contains valuable information for guiding product design, such as customer preferences, market demands, product evaluation, and visual display: online product reviews reflect customer evaluations and requirements; product images contain information of shape,color, and texture which can inspire designers to get initial design schemes more quickly or even directly generate new product images. How to efficiently collect product design related data and exploit them effectively during the whole product design process is thus critical to modern product design. This paper aims to conduct a comprehensive survey on big data driven product design. It will help researchers and practitioners to comprehend the latest development of relevant studies and applications centered on how big data can be processed, analyzed, and exploited in aiding product design. We first introduce several representative traditional product design methods and highlight their limitations. Then we discuss current and potential applications of textual data, image data, audio data, and video data in product design cycles. Finally, major deficiencies of existing data driven product design studies and future research directions are summarized. We believe that this study can draw increasing attention to modern data driven product design.
\end{abstract}


\section{Introduction}
\label{sec:Introduction}
As the materialization of consumer requirements, products are the foundation for enterprise development. Successful products can improve user satisfaction, stimulate purchase desire, increase sales, and promote the completion of 'product-money-new product', so that enterprises can survive in the fierce market competition\cite{RN263}. In recent decades, with the advancement of manufacturing and technology, product design has undergone major changes in design concepts and methods. The design concept is changing from 'form follows function', 'technology first', and 'product-centric' to 'form follows emotion' and 'user-centric' \cite{RN264}. The design method is changing from experiential design and fuzzy design to intelligent design, computer-aided design, and multi-domain joint design \cite{RN167, RN265, RN151}. At the same time, customers pay more and more attention to the spiritual and emotional requirements, which are diversified and personalized, especially in product appearance \cite{RN40, RN268}. Therefore, this article focuses on product appearance (the product design in the following text is for appearance), involving outlook and function.

At present, the research of traditional product design methods has been very mature, such as the theory of inventive problem solving (TRIZ), quality function deployment (QFD), and axiomatic design (AD) \cite{RN13, RN12, RN17, RN16, RN38}. However, difficulty in capturing user requirements, lack of visual display, and product evaluation are still significant difficulties in product design \cite{RN140, RN88, RN8}. The foundation of capture user requirement and product evaluation is data acquisition. Usually, traditional methods are implemented in two ways. One is the manual survey, which is time-consuming and labor-intensive, and data distribution may be uneven \cite{RN43}. Furthermore, it is one-time and cannot be updated, which is an obstacle in the fast era \cite{RN1}. The other is to rely on designers' or experts' experience and intuition, which is highly subjective, uninterpretable, and risky \cite{RN41}. There are other drawbacks to the traditional method, which we will discuss in detail in Section 3. These deficiencies have led to long product development cycles, poor predictability, and low success rates, which are essential to enterprises. Recently, some studies have found that big data can alleviate these problems to a certain extent in an automated and intelligent way \cite{RN1, RN114}. The "Made in China 2025" plan also encourages the use of big data and Internet+ to enhance product innovation capabilities. Big data has become a popular focus in the field of product design \cite{RN6, RN5}.

\begin{figure}
  \centering
  \includegraphics[width=15cm]{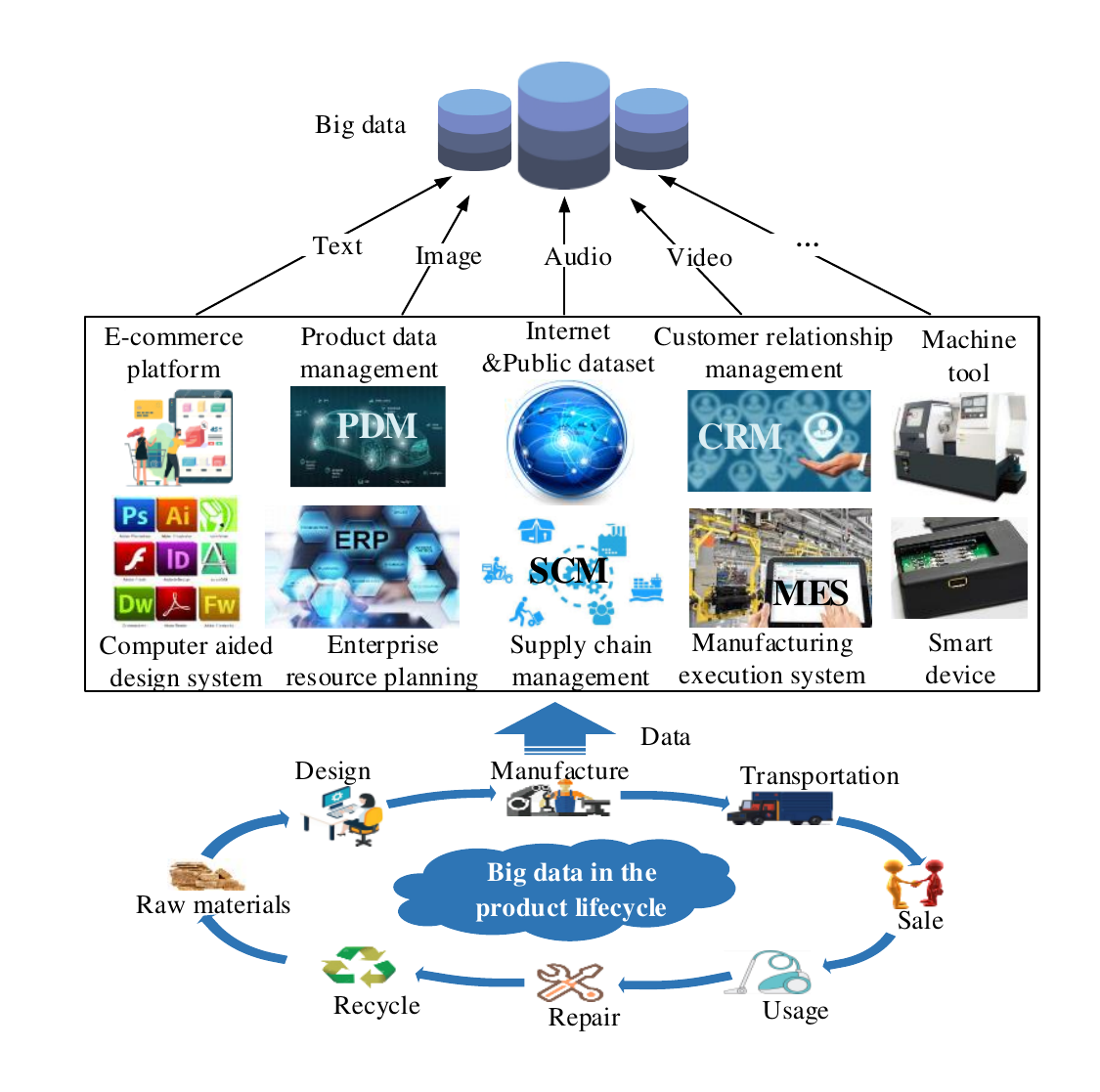}
  \caption{Big data in the product lifecycle. The product lifecycle is the entire process from product development to scrap, including design, manufacture, transportation, sale, usage, repair, and recycle.}
  \label{fig:fig1}
\end{figure}

With the widespread application of information technologies such as the Internet of things, the Internet, and automation, a large amount of data has accumulated in the product lifecycle (Figure \ref{fig:fig1}), which is growing exponentially every day \cite{RN7, RN8}. By reviewing a large amount of literature, Zhang et al. \cite{RN10} concluded that the data in the product lifecycle is very useful, and the hidden knowledge can be obtained through appropriate analysis methods. The product lifecycle contain a lot of product feedback information, such as user preference, market demand, and visual display \cite{RN11, RN9, RN269}. This information is valuable for guiding product design and has aroused increasing interest in product design and big data fields. On the one hand, processing and analyzing so much data will result in new challenges that need to be appropriately addressed. On the other hand, a successful analysis will bring better products. How to obtain valuable information from big data and applied to design is the difficulty and focal point of existing research \cite{RN114, RN11, RN9}.

\begin{figure}
  \centering
  \includegraphics[width=15cm]{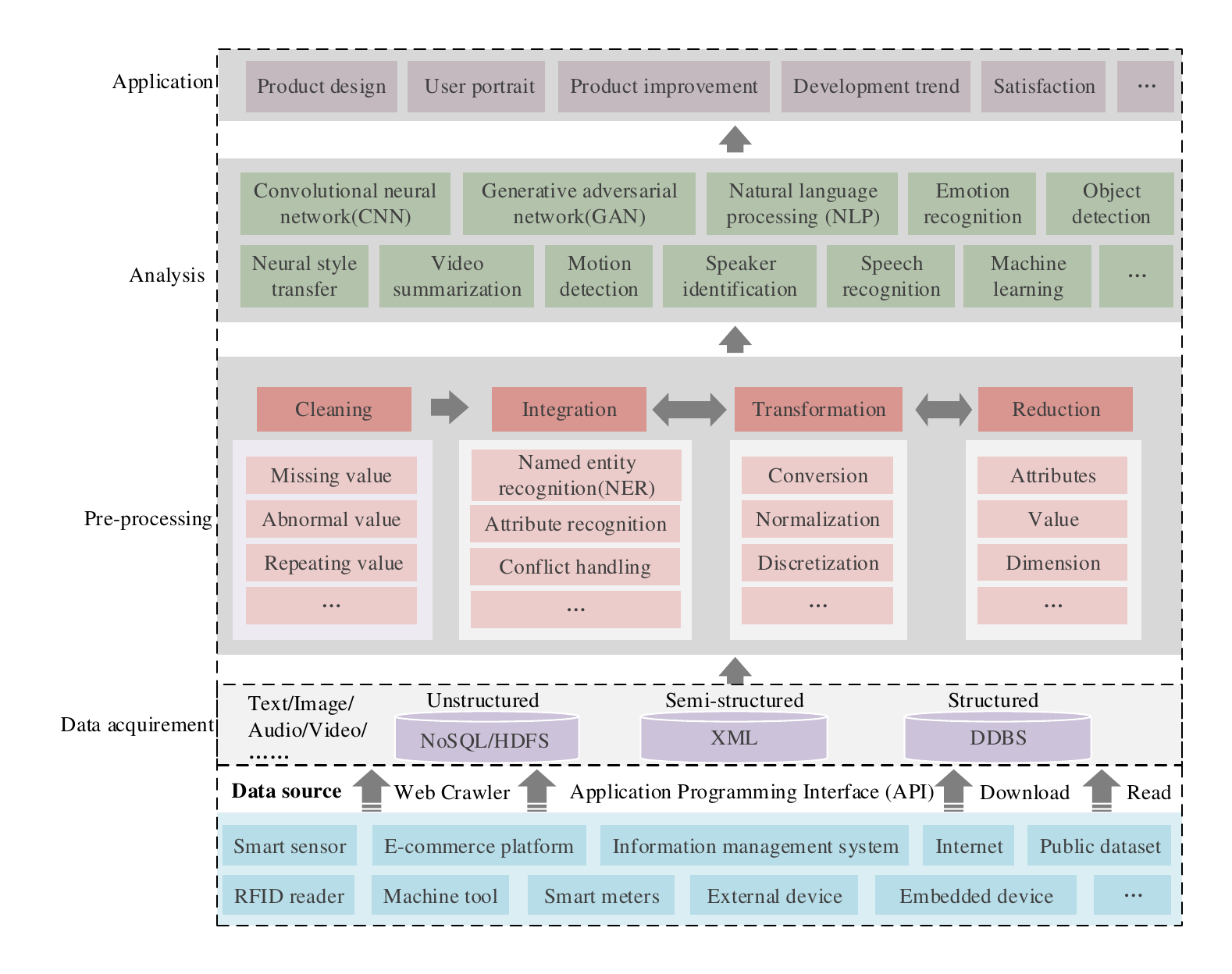}
  \caption{A framework for mining product-related information from big data. The framework includes data acquisition, pre-processing, analysis, and application. Data acquisition is to collect data from data sources; pre-processing aims to standardize and standardize the acquired data; data analysis reveals the hidden knowledge and information; application directly reflects big data value.}
  \label{fig:fig2}
\end{figure}

Big data in the product lifecycle is characterized by multiple data types, large volume, low-value density, and fast update, so it is not easy to handle with conventional techniques and algorithms. Emerging technologies that have the potential to process big data include convolutional neural network (CNN), generative adversarial network (GAN), natural language processing (NLP), neural style transfer, motion detection, speech recognition, video summarization, and emotion recognition. In Figure \ref{fig:fig2}, an overview of mining product-related information from big data is presented. In this paper, we mainly focus on data processing and application.

\begin{figure}
  \centering
  \includegraphics[width=15cm]{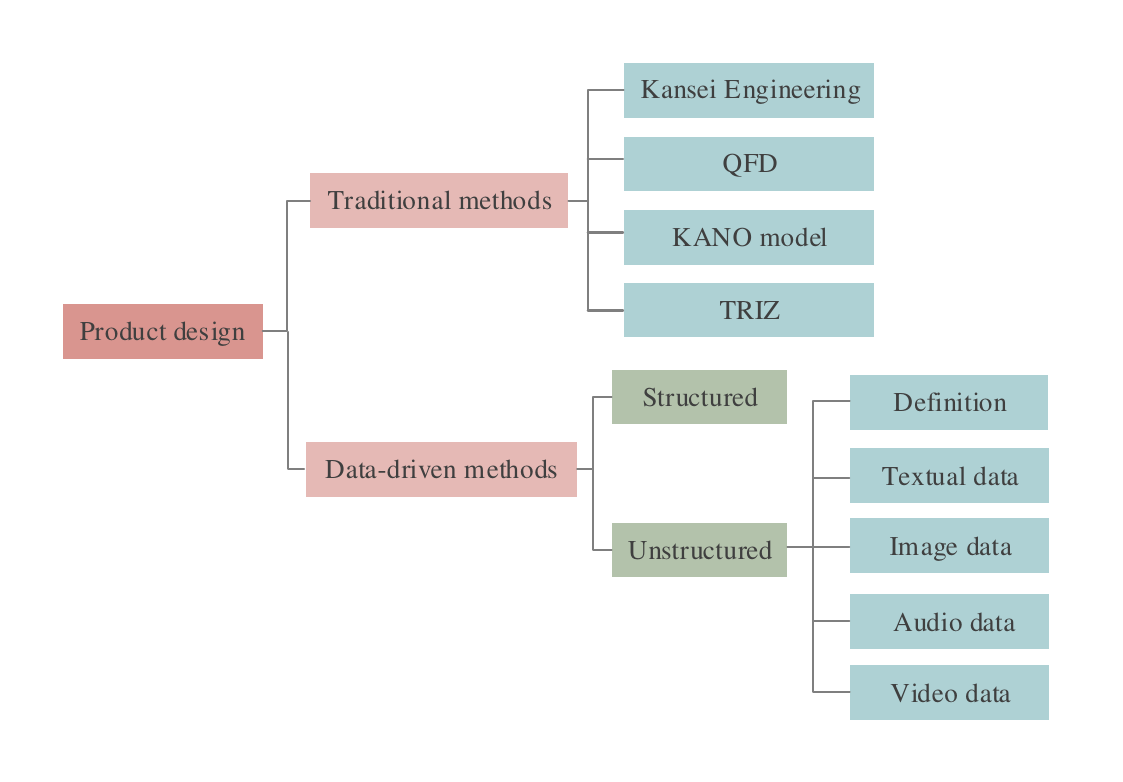}
  \caption{The research framework. Through Kansei Engineering, QFD, and KANO model, we revealed the shortcomings of traditional product design methods and showed that big data could improve them.}
  \label{fig:fig3}
\end{figure}

In recent years, product design based on big data has become a research hotspot. However, to our best knowledge, before this work, the applications of big data in product design have not been summarized. Therefore, a comprehensive literature review is required to provide theoretical foundations that can be adopted to develop scientific insights in this area, and to help enterprises to shorten the time of the product development cycle, save costs, improve user satisfaction, and promote the development of product design toward automation and intelligent. In this paper, we introduced the traditional product design method and the big data based method, summarized their shortcomings and difficulties, and focus on the application, technology, and process flow of structured data, textual data, image data, audio data, and video data. The research framework is shown in Figure \ref{fig:fig3}.

Figure \ref{fig:fig3} shows the structure of the paper. In Section \ref{sec:section2}, we introduce several key tasks in product design. In Section \ref{sec:Traditional}, several representative traditional product design methods are introduced, and their shortcomings are summarized. In Section \ref{sec:Product}, we provide a comprehensive review focuses on how big data are being processed and exploited to obtain helpful information for product design. Moreover, some shortcomings and potential future studies are presented in Section \ref{sec:Product}. In Section \ref{sec:Conclusions}, we summarize our work, and highlight the development direction of the product design field.

\section{Key tasks in product design}
\label{sec:section2}

\begin{figure}
  \centering
  \includegraphics[width=15cm]{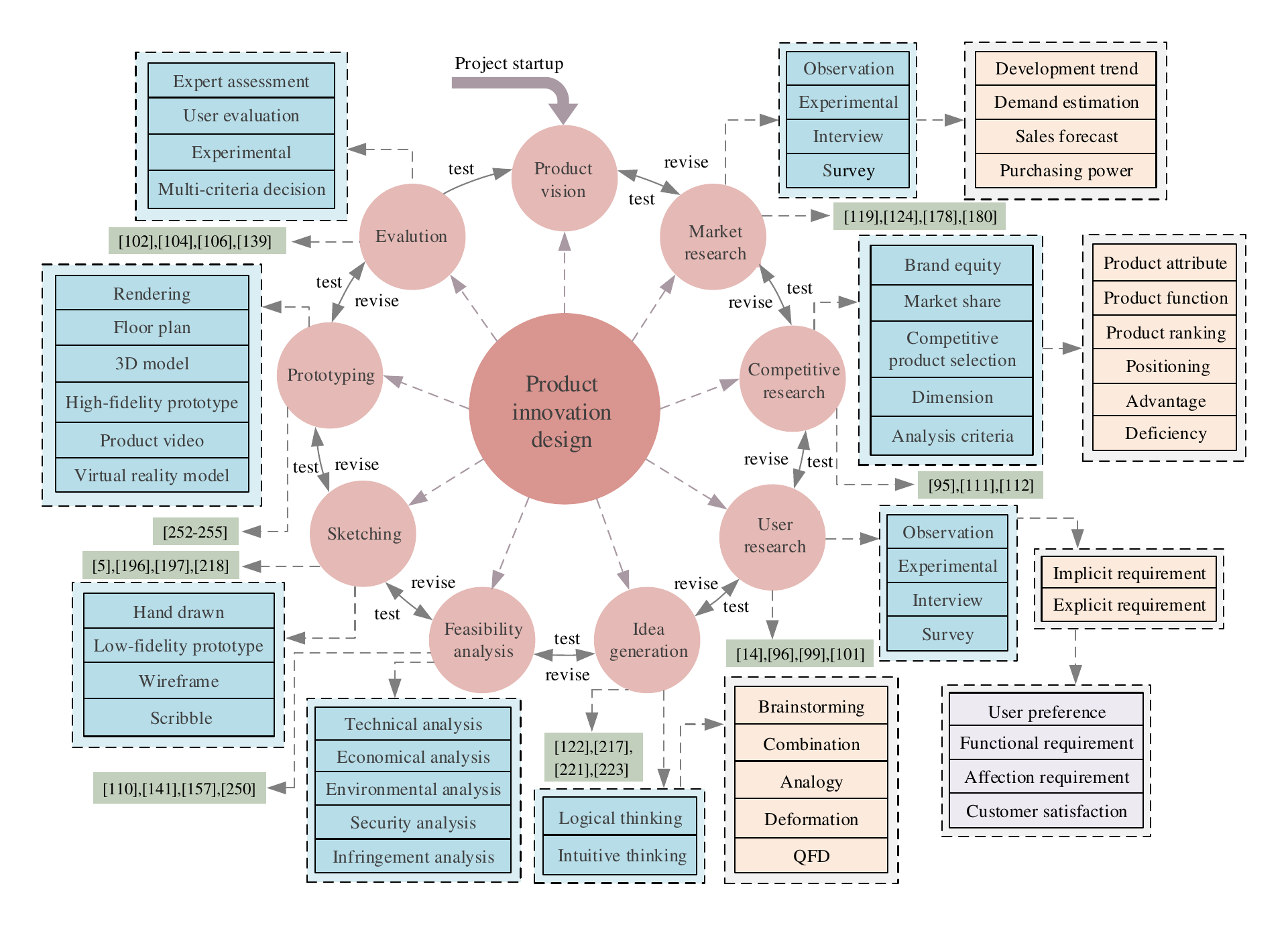}
  \caption{The product design process. The dotted line represents the affiliation, and the solid line represents the product design process. Major references of data driven methods are given. The user research, idea generation, and product display (sketching and prototyping) are the key tasks in product design.}
  \label{fig:fig4_new}
\end{figure}

The product design process can be defined as a series of steps that are followed to develop a design scheme. It’s impossible to provide a universal design process that fits all products, more or less adjustments are needed in a specific project. By consulting a large number of articles, we summarized a typical product design process, as shown in Figure \ref{fig:fig4_new}. It includes nine design tasks which are namely product vision definition, market research, competitive research, user research, idea generation, feasibility analysis, sketching, prototyping, and scheme evaluation. The product vision definition is actually done before the design process even starts, and it helps the product team clarify the overall goal. Market research aims to understand the development trend, the user’s demand and purchasing power. Competitive analysis compares existing products from multiple dimensions and derives their advantages and deficiencies. 

\textbf{User research} is one of the key tasks of product design, which is supported by data. Data are collected through observation, experiment, interview, and survey. In terms of the quality and quantity of data collected, traditional methods are far inferior to big data driven methods. To extract user requirements, preferences, and satisfaction, traditional methods usually adopt Kansei Engineering, QFD, KANO model, AD, and affective design \cite{RN17,RN13, RN39, RN57}; while in the big data driven methods, technologies such as natural language processing, speech recognition, emotion recognition, and intelligent video analysis are widely used. Depending on different object-oriented, product design methods can be divided into product-centric and user-centric. The product-centric approach focuses on improving performance and emphasizes that users passively adapt to the product; the user-centric approach concerns the satisfaction of users' spirit requirements and emphasizes that products actively adapt to users. With the improvement of human living standards, users' requirements have become more and more diversified and personalized, prompting the user-centric approach to become the mainstream. Kansei Engineering, QFD, KANO model, and AD are typical user-centric methods \cite{RN17,RN13, RN39, RN57}.

\textbf{Idea generation} is also one of the key tasks of product design. According to different innovative thinking, idea generation methods can be divided into two types: logical thinking and intuitive thinking \cite{RN2}. The former focuses on detailed analysis and decomposition of problems, including TRIZ, universal design, and AD  \cite{RN13, RN12, RN38}. The latter aims to inspire designers, including brainstorming, bionics, analogy, combination, and deformation \cite{RN15, RN14, RN16}. Nathalie and John \cite{bonnardel2020brainstorming} adopted brainstorming to favor ideation during design activities. Youn et al. \cite{youn2015invention} found that the combination is the major driver of invention by analyzing a large number of patents. Lai et al. \cite{lai2006user} obtained new products by combining different product form and color. Zarraonandia et al. \cite{zarraonandia2017using} used the combinatorial creativity to digital game design. The next key tasks of product design is feasibility analysis, which involves technical analysis, economical analysis, security analysis, infringement analysis, and environmental analysis. It is worth noting that technical analysis will explore whether there are contradictions among various technical attributes and resolve them.

\textbf{Product display} allow users, designers, and experts to intuitively understand the designed product. Sketches and prototypes are visualizations of ideas. Traditional methods require designers to process drawing skills, such as hand drawing, or 3D modeling. The big data driven method provides simpler operations and does not require drawing skills. In addition, the traditional display way are mostly images or 3D models, while the big data driven method can be video or interaction, which will be more intuitive and interesting.

\textbf{Evaluation} is the last key task of product design. The designed product will be returned to users, experts, designers, or decision makers for feedback. Improve the product based on the feedback and verify that whether the final product is consistent with the vision. User evaluation can also be considered as part of user research task. Mattias and Tomohiko \cite{2012A} proposed a new evaluation method, which is achieved based on the importance of various customer value and each offering’s contribution to the value as well as the customer’s budget. They applied this method to a real life case at an investment machine manufacturer.

\section{Traditional product design methods}
\label{sec:Traditional}

\subsection{Kansei Engineering}

\begin{figure}
  \centering
  \includegraphics[width=13cm]{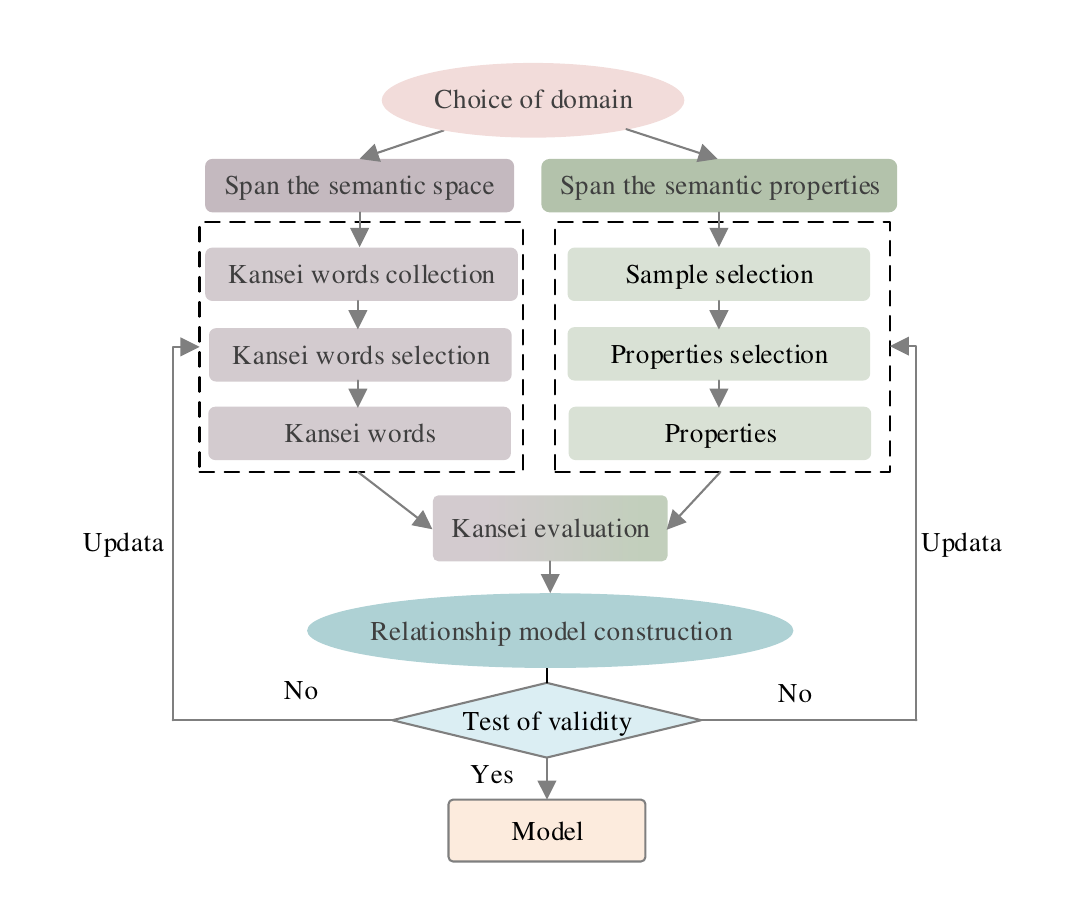}
  \caption{The framework of Kansei Engineering.}
  \label{fig:fig4}
\end{figure}

Kansei Engineering is the most widely used user-centric method \cite{RN18}, and is usually used in the user research task. In the 1970s, Nagamechi pointed out that in Japan, which already had sufficient material wealth, the consumption trend tilted from function to sensibility, and sensibility became the core of product design \cite{RN19}. Nissan, Mitsubishi, Honda, and Mazda used Kansei Engineering to improve car positioning, shape, color, dashboard, etc., which brought great success to the Japanese automobile industry \cite{RN20, RN22, RN21}. In the late 1990s, Kansei Engineering spread to Europe. To make Kansei Engineering applicable to European culture, Schütte proposed a modification strategy to reduce the complexity \cite{RN23, RN25, RN26, RN24}. Besides, Schütte also discussed the selection of samples \cite{RN27} and visualized Kansei Engineering steps \cite{RN28}. Nagamechi laid the foundation of Kansei Engineering and never stopped researching. In the latest research \cite{RN29}, he encouraged scholars to introduce artificial intelligence into Kansei Engineering. Figure \ref{fig:fig4} shows the framework of Kansei Engineering.

The critical process of Kansei Engineering includes the collection of Kansei words, the selection of Kansei words, the Kansei evaluation, and the relationship model construction. (i) The collection of Kansei words. Conventional collection sources include magazines, documents, manuals, experts, interviewees, product catalogs, and e-mails. The collected words may come from different fields and have poor pertinence and applicability. (ii) The selection of Kansei words. Some studies rely on manual screening, which brought  intense subjectivity to the results; some used principal component analysis, factor analysis, hierarchical clustering analysis, and K-means to help selection. Considering the high frequency of important Kansei words, Djatna et al. \cite{RN30} adopted the Term Frequency-Inverse Document Frequency (TF-IDF) to select Kansei words and completed the packaging design of the tea powder product. Shieh and Yeh \cite{RN31} collected 100 sets of Kansei words through experts, web searches, sports magazines and periodicals, and advertisements, and then used cluster analysis chose four sets of Kansei words. (iii) Kansei evaluation. Most studies use a semantic differential scale or Likert scale to design questionnaires and obtain Kansei evaluations from survey results. Besides, a few scholars pay attention to physiological response \cite{RN33, RN32}. They believe that the physiological signal is more reliable than the defined score. Kemal et al. \cite{RN34} used the eye-tracker to obtain the area of interest (AOI), scan path, and heat maps, then evaluated the ship's appearance based on these objective data. A single physiological signal is one-sided. Therefore. Xiao and Cheng \cite{RN35} designed several experiments involving eye-tracking, skin conductance, heart rate, and electroencephalography. We notice that physiological signals are more about Kansei's intensity than content. (iv) Relationship model construction. Connecting user affections to the product parameters is the final step. The relationship model can be constructed by quantification theory, support vector machine, neural network, etc. \cite{RN36}

Although Kansei Engineering has achieved successful applications, there are still several deficiencies: (i) the collected Kansei words are inadequate in pertinence and number; (ii) the small number of samples and subjects; (iii) both questionnaires and physiological signals are susceptible to time, environment, and concentrated subjects, which will reduce the authenticity and objectivity; (iv) limited by life background and work experience, subjects' evaluations in the questionnaire vary greatly; (v) the survey scope is small, which may cause uneven data distribution. These deficiencies are mainly in data acquisition, especially the collection of Kansei words and Kansei evaluation. The quantity and quality of survey data will directly affect the training of the relational model. Therefore, data acquisition needs to be solved urgently.

 \subsection{KANO model}

\begin{figure}
  \centering
  \includegraphics[width=15cm]{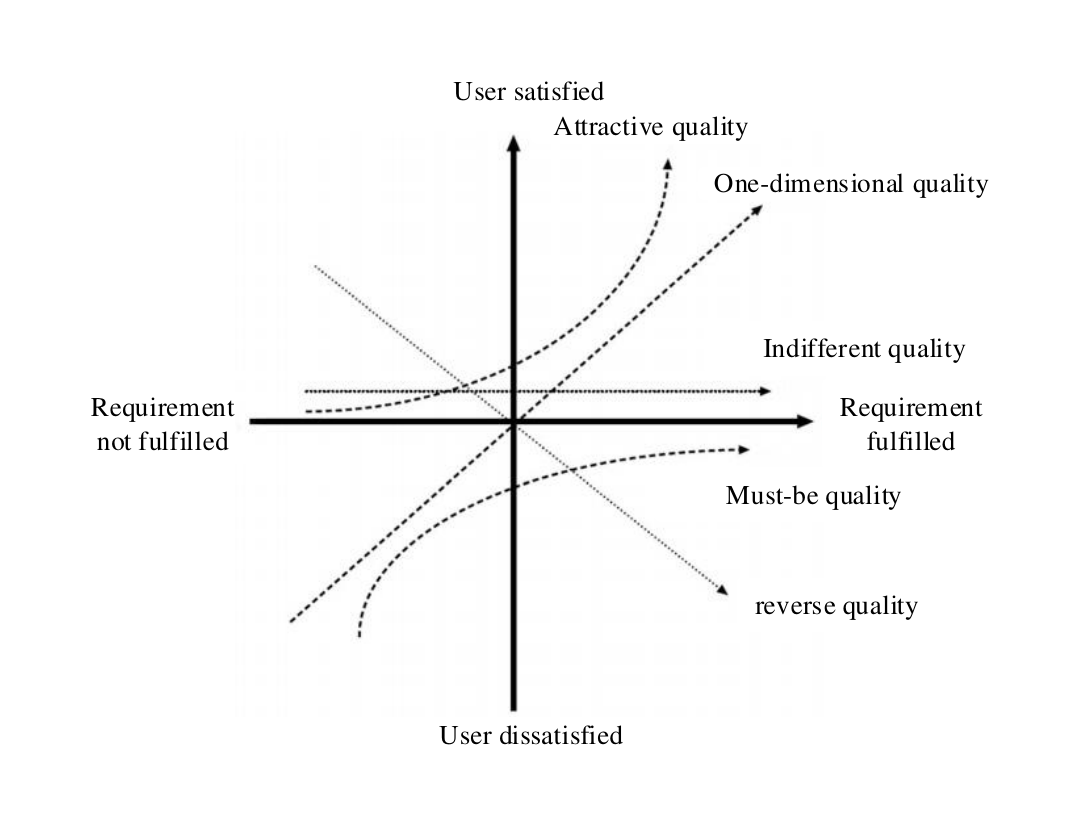}
  \caption{The KANO diagram. The horizontal axis indicates how fulfilled the requirement is, and the vertical axis indicates how satisfied the user is.}
  \label{fig:fig6}
\end{figure}

The Kano model is used for user research task, and its purpose is to classify user requirements. For a long time, researchers roughly think that user satisfaction is directly proportional to product performance. However, fulfilling the individual product performance to a great extent does not necessarily lead to high user satisfaction, and not all performances are equally important. In the 1980s, Noriaki Kano conducted a detailed study on user satisfaction and proposed the KANO model \cite{RN55}. The KANO model shows the nonlinear relationship between user satisfaction and requirements fulfillment (Figure \ref{fig:fig6}). The user requirements are divided into five types: must-be quality, one-dimensional quality, attractive quality, indifferent quality, and reverse quality \cite{RN56, RN57}.

Must-be quality refers to requirements that users took for granted. If fulfilled, users are just neutral, but if not fulfilled, users are incredibly dissatisfied. For example, the call function for the mobile phone. One-dimensional quality refers to requirements proportional to user satisfaction, such as the battery for the mobile phone. Attractive quality will increase satisfaction when fulfilled and not cause dissatisfaction when not fulfilled, for example, the temperature display for the cup. Indifferent quality has no impact on user satisfaction, for example, the built-in pocket for the coat. Reverse quality will result in dissatisfaction when fulfilled, for example, the pearl for the men's wear. Yao et al. \cite{RN58} categorized twelve features of mobile security applications (MSA) through a structured KANO questionnaire and discussed MSA's optimization from the perspective of enterprises and customers. Avikal et al. \cite{RN59} integrated the KANO model with QFD to examine customer satisfaction based on aesthetic sentiments.

According to the priority assigned by the KANO model, the enterprise must just make the must-be quality reach the threshold, and more investment is a waste; invest in one-dimensional quality as much as possible; under the rich conditions, grasp the attractive quality; avoid the reverse quality; never waste resources on the indifferent quality. All customer requirements are not equal, and even they are in the same category. Since the KANO model cannot distinguish the differences among requirements within the same category \cite{RN60}, Lina et al. \cite{RN61} proposed the IF-KANO model, which adopted logical KANO classification criteria to categorize requirements. They take elevators as an example, and results showed that both load capacity and operation stability belong to one-dimensional quality, but the priority of operation stability is slightly higher than load capacity.

The KANO questionnaire is the data acquisition of the KANO model, and the conventional KANO questionnaire only provides a single option or options within a given range, which does not fully consider the ambiguity and complexity of customers. Therefore, many scholars have put efforts into improving the KANO questionnaire in terms of options \cite{RN64, RN62}. Mazaher et al. \cite{RN65} proposed the fuzzy Kano questionnaire, which replaces options with percentages and allows subjects to make multiple choices. Chen and Chuang \cite{RN67} introduced the Likert scale to identify the degree of satisfaction or dissatisfaction. Cigdem and Amitava \cite{RN68} integrated the Servqual scale and the KANO model in a complementary way. A few scholars noticed that text could express opinions better than scale, but the text is difficult for traditional statistical methods \cite{RN262}. In addition to work on questionnaire options, some studies focused on questioning skills to improve the KANO questionnaire. Bellandi et al. \cite{RN69} pointed out that questions should avoid polar wording.

Although the KANO questionnaire has been improved, as mentioned in section 3.1, the questionnaire has its limitations. Furthermore, the shortcomings of the Kano model are as follows: (i) requirements are given in advance, lack of acquisition research; (ii) the KANO model only focuses on the category and rating of customer requirements, and its application is primitive.

\subsection{QFD}

\begin{figure}
  \centering
  \includegraphics[width=12cm]{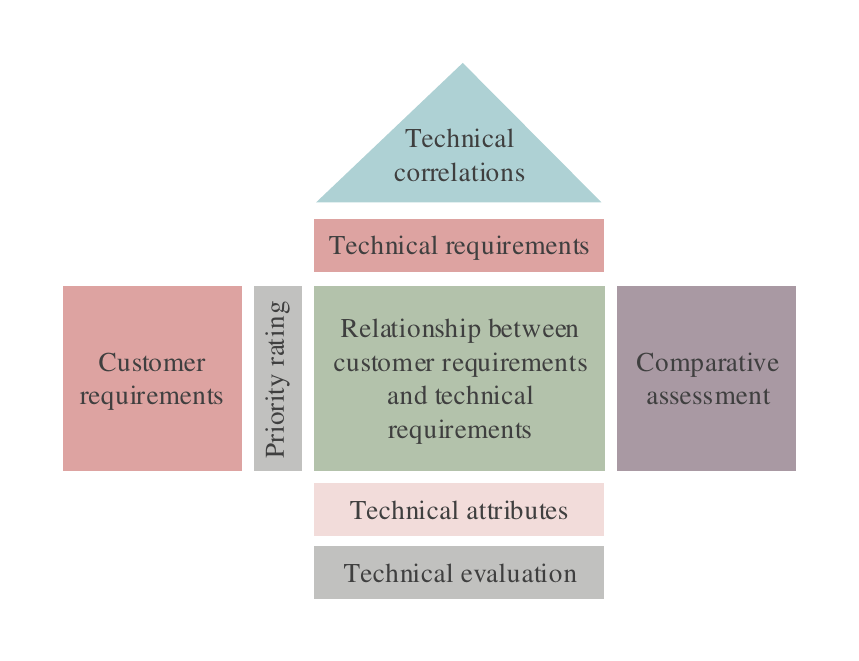}
  \caption{The house of quality (HoQ). HoQ consists of customer requirements, priority rating, technical requirements, technical correlations, relationship matrix, comparative assessment, technical attributes, and technical evaluation.}
  \label{fig:fig5}
\end{figure}

QFD was proposed by Japanese scholars Yoji Akao and Shigeru Mizuno in the 1970s, aiming to transform customer requirements into technical attributes, parts characteristics, key process operations, and production requirements \cite{RN37, RN39, RN38}. The house of quality (HoQ) is the core of QFD, shown in Figure \ref{fig:fig5}. QFD can be used for idea generation, competitive analysis, and technical analysis tasks.

Starting from meeting market demands and driven by customer requirements, QFD explicitly transforms requirements information into specific information directly used by design, production, and sales departments, to ensure that the products can meet customer requirements \cite{RN40}. Tomohiko \cite{RN41} used QFD to decompose customer requirements layer by layer, identify the corresponding characteristics, and build an HoQ to guide hair dryers' design. Moreover, the customer requirement from various stakeholders such as recyclers, production engineers, and customers. Yan et al. \cite{RN42} applied QFD to competitive analysis and obtained product improvement strategies. Noting the inherently vague and ambiguity of customer requirements \cite{RN44, RN43}, Cengiz et al. \cite{RN45} proposed fuzzy QFD. Considering that the technical characteristics in the HoQ may be contradictory, Wang et al. \cite{RN46} combined QFD with TRIZ and used the contradiction matrix to obtain the solution. When ranking the importance of customer requirements, QFD usually relies on experts, which leads to intense subjectivity. To solve this problem, scholars have introduced the multi-criteria decision-making (MCDM) methods \cite{RN49, RN47, RN48, RN50}. Similar to Kansei Engineering, QFD also uses a definite score to express evaluation, ignoring the uncertainty of user and expert scoring. Therefore, some studies introduced the rough theory, interval-valued fuzzy-rough sets, and grey relational analysis \cite{RN51, RN52, RN53, RN54}. 

Despite some progress achieved by QFD, limitations exist: (i) The relationship between customer requirements and characteristics is judged by experts manually, which heavily relies on experts; (ii) The evaluation of satisfaction is completed by a small number of subjects; (iii) Customer requirements are directly given by experts, customers, designer, or summarized by scholars, lacking the data acquisition. 

 \subsection{TRIZ}

TRIZ is an approach of finding inventive solutions by identifying and resolving contradictions. It can be used for idea generation as well as technical analysis tasks. TRIZ was developed by Soviet inventor Genrich Altshuller and his colleagues, beginning in 1946 \cite{moussa2017reviewing}. Altshuller analyzed over 400,000 invention patents and developed the technical contradiction, the concept of ideality of a system, contradiction matrix, 40 principles of invention (Table \ref{tab:table1_new}), and 39 engineering parameters (Table \ref{tab:table2_new}). Altshuller also observed smart and creative people, uncovered patterns in their thinking, and developed thinking tools and techniques to model this "talented thinking".

\begin{table}
 \caption{The 40 inventive principles of TRIZ.}
  \centering
  \begin{tabular}{ p{1cm} p{5cm} p{1cm} p{5cm}}
    \toprule
    No.&Inventive principle&No.&Inventive principle \\
    \midrule
    1&Segmentation&21&Skipping\\
    2&Taking out&22&Blessing in disguise\\
    3&Local quality&23&Feedback\\
    4&Asymmetry&24&Intermediary\\
    5&Merging&25&Self-service\\
    6&Universabrationlity&26&Copying\\
    7&Nested dol1&27&Cheap short-living\\
    8&Anti-weight&28&Mechanics substitution\\
    9&Preliminary anti-action&29&Pneumatics and hydraulics\\
    10&Preliminary action&30&Flexible shells and thin films\\
    11&Beforehand cushioning&31&Porous materials\\
    12&Equipotentiality&32&Colours changes\\
    13&The other way around&33&Homogeneity\\
    14&Spheroidality&34&Discarding and recovering\\
    15&Dynamics&35&Parameter changes\\
    16&Partial or excessive actions&36&Phase transitions\\
    17&Another dimension&37&Thermal expansion\\
    18&Mechanical vibration&38&Strong oxidants\\
    19&Periodic action&39&Inert atmosphere\\
    20& Continuity of useful action&40&Composite material film\\
    \bottomrule
  \end{tabular}
  \label{tab:table1_new}
\end{table}

\begin{table}
 \caption{The 39 engineering parameters.}
  \centering
  \begin{tabular}{ p{1cm} p{5cm} p{1cm} p{5cm}}
    \toprule
    No.&Engineering parameter&No.&Engineering parameter\\
    \midrule
    1&Weight of moving object&21&Power\\
    2&Weight of nonmoving object&22&Waste of energy\\
    3&Length of  moving object&23&Waste of substance\\
    4&Length of nonmoving object&24&Loss of information\\
    5&Area of  moving object&25&Waste of time\\
    6&Area of nonmoving object&26&Amount of substance\\
    7&Volume of moving object&27&Reliability\\
    8&Volume of nonmoving object&28&Accuracy of measurement\\
    9&Speed&29&Accuracy of manufacturing\\
    10&Force&30&Harmful factors acting on object\\
    11&Tension, pressure&31&Harmful side effects\\
    12&Shape&32&Manufacturability\\
    13&Stability of object&33&Convenience of use\\
    14&Strength&34&Reparability\\
    15&Durability of moving object&35&Adaptability\\
    16&Durability of nonmoving object&36&Complexity of device\\
    17&Temperature&37&Complexity of control\\
    18&Brightness&38&Level of automation\\
    19&Energy spent by moving object&39&Productivity\\
    20&Energy spent by nonmoving object\\
    \bottomrule
  \end{tabular}
  \label{tab:table2_new}
\end{table}

The use of TRIZ follows four steps: (i) Define your specific problem. (ii) Abstraction-converting your specific problem into general problem. (iii) Mapping-finding solution for solving general problem. (iv) Concretizing-projecting TRIZ general solution to develop your specific solution. Wang \cite{RN16} used principle 1 (segmentation), principle 5 (combining) and principle 28 (replacement of a mechanical system) in Table \ref{tab:table1_new} to design phone-cameras. Yamashina et al. \cite{RN38} integrated QFD and TRIZ method to perform washing machine design. Ai et al. \cite {ai2020low} designed low-carbon products from the perspectives of technical system and human use, and obtained measures to improve energy efficiency through TRIZ.

\subsection{Summary of limitations}
Among the user-centered methods mentioned above, the QFD focuses on constructing the HoQ to map customer requirements to product parameters. The KANO model only emphasizes the classification and rating of customer requirements. Both the KANO model and QFD adopt the pre-given requirements directly, and lack of acquisition research. The Kansei Engineering focuses on the acquisition, expression, and mapping of customer requirements, and it guides product design by building the relationship model between customer requirements and product properties. Although Kansei Engineering is a comprehensive method, it still has shortcomings in the data collection process. In order to obtain the customers' affective responses on different products, many methods are widely used in traditional practices, such as user interviews, questionnaires, focus groups \cite{RN88}, etc. Here we summarized their disadvantages and advantages in Table \ref{tab:table1}.

\begin{table}
 \caption{Comparison of user requirement mining methods.}
  \centering
  \begin{tabular}{ p{3.5cm} p{3.5cm} p{3.5cm} p{3.5cm}}
    \toprule
    Methods&Description&Advantages&Disadvantages \\
    \midrule
    User interview&The interviewer talks directly with the subject.& Detailed; easy to implement &Time-consuming; one-time; subjective; labor-intensive; small survey scope\\
    Questionnaire&Record subjects' opinions on specific questions.&Easy to implement&Subjective; one-time; time-consuming; labor-intensive; undispersed region and time; small survey scope\\
    Web-based questionnaire&Distribute questionnaires on the Internet.&Dispersed region and time; large survey scope&Subjective; one-time\\
    Focus-group&Observe the opinions and behaviors of a group on the subject.&Real data; low cost; in-depth questions&Time-consuming; labor-intensive; one-time; complex; small survey scope\\
    Usability testing&Subjects test the product and give feedback.&Detailed; high reliability&One-time; labor-intensive; small survey scope; time-consuming\\
    Experience&Analyze the data generated by consumers during usage.&Easy to implement&Time-consuming; slow; inefficiency; subjective; small survey scope\\
    Experimental&Record the psychological and physiological data of the subject.&Detailed; high reliability& Expensive; one-time; time-consuming; labor-intensive; complex; small survey scope\\
    \bottomrule
  \end{tabular}
  \label{tab:table1}
\end{table}

In general, the disadvantages of traditional product design methods include: (i) Difficulty to capture accurate customer requirements. With the increasing diversification and complexity of customer requirements, it is more and more difficult for enterprises to determine product positioning. (ii) Various shortcomings in the process of data acquisition, such as: manual, time-consuming, laborious, hard to update, subject to time and place, and small survey scope. (iii) Small sample size, poor real-time data, and small amount of data, 
while the quantity and quality of samples and data are directly related to product development direction. In addition, traditional data collection takes a lot of time, which makes the data quickly outdated. (iv) The subjects are susceptible to time and environment, which will directly affect the survey results. Moreover, the data is generated by subjects, and there are differences between them and real users. (v) Heavy dependence on expert knowledge. It will increase the workload and elongate the time of the product development cycle. Moreover, different experts have different backgrounds and experiences, which brings uncertainty and subjectivity to designs. (vi) Lacking the intuitive approach to inspiring designers. Visualization has a crucial influence on inspiration. (vii) In the early design stage, design schemes' lack of visual display is not conducive to enterprises to hold product development direction. Deciders only rely on imagination and cannot see designs with their own eyes, which will lead to a high probability of failure. Failed products may waste many resources and even make the enterprise bankrupt. These deficiencies mentioned above have clearly stagnated the development of product design.

The advent of the big data era provides ideas and technologies for addressing the shortcomings of traditional product design methods and enhancing innovation capabilities. Big data and its analysis technology can reduce subjectivity, expand survey scope, process data automatically (including data acquisition, update and analysis), accurately acquire user requirements, and provide intuitive visual presentations. We take the product evaluation as an example to illustrate the survey scope and data processing. By taking the world’s e-commerce platforms, social media, and review websites as data sources, we can use web crawlers to obtain customer reviews of products, and adapt NLP technology to automatically extract information such as product attributes, opinion words, and sentiment orientations, etc., to obtain customers' evaluations on products around the world. In addition, since customer evaluations are dynamic, we only need to add a piece of updated code to get real-time evaluation. In next section, we will introduce a variety of new-generation data driven product design methods, which are the focus of this article.

  \section{Product design based on big data}
  \label{sec:Product}
  \subsection{Definition of big data}

With the rapid development of the Internet, the Internet of Things (IoT), and communication technologies, the amount of data is expanding exponentially \cite{RN71, RN72}. We have entered the era of big data. In the 1980s, Alvin Toffler proposed the term "big data" in \emph{The Third Wave} \cite{toffler1980third}; he enthusiastically praised big data as the third wave’s cadenza \cite{RN73}. However, there is no unified definition of big data. International Data Corporation (IDC) defined big data as: "big data technologies describe a new generation of technologies and architectures, designed to economically extract value from very large volumes of a wide variety of data, by enabling high-velocity capture, discovery, and/or analysis." Gartner defined big data as: "Big data is high-volume, high-velocity and high-variety information assets that demand cost-effective, innovative forms of information processing for enhanced insight and decision making." Mckinsey defined big data as: "datasets whose size is beyond the ability of typical database software tools to capture, store, manage, and analyze." Although these popular definitions are derived from different perspectives, they all reflect the 5V properties (Volume, Velocity, Variety, Value, and Veracity) of big data. The same is true for big data in the product lifecycle.

 \textbf{\emph{Volume }}The data sources of product lifecycle include (i) equipment. Such as sensors, machine tools, RFID readers, and smart mobile devices. (ii) Management system. Such as manufacturing execution system (MES), enterprise resource planning (ERP), customer relationship management (CRM), supply chain management (SCM), and product data management (PDM). (iii) Internet. Browse, click, purchase, score, commenting, etc., on e-commerce platforms or social networking platforms. Moreover, Internet data also include public data sets.
 
 \textbf{\emph{Veracity }}Data in the product lifecycle is actually generated, some of which come directly from the user. For example, online reviews on e-commerce platforms can only be posted by users who have made a purchase. Although there are some spam reviews, they can be automatically removed through analysis techniques \cite{RN76, RN74, RN75}. High-quality data is the foundation of the product design. Furthermore, user-generated data is valuable for capturing user requirements and product evaluations.
 
 \textbf{\emph{Velocity }}There are multiple data sources in the entire lifecycle of a product, and these data are generated quickly. Considering that user requirements and evaluations are dynamic rather than immutable, updated feedback can better help companies capture user requirements and promptly respond to them.

 \textbf{\emph{Value }}The value of big data depends on the role of data analysis results in applications. Big data in the product lifecycle can expose users' hidden behavioral patterns and even shed light on their intentions. How to obtain this information from massive data and guide design has been a crucial problem for the product design field.

 \textbf{\emph{Variety }}Many types of data are generated during a product’s lifecycle, including voice, video, images, text, location, and numbers. They have different aids in the product design process. For example, the text is helpful for user requirements, while images are useful for visualization and inspire designers.

 Big data in the product lifecycle can be divided into structured data, semi-structured data, and unstructured data. The structure and content of structured data are separate, while the structure and content of semi-structured data are mixed. Unstructured data have no fixed structure, including audio, video, images, text, and location.

 \subsection{Product design based on structured data}
 Structured data has been formatted and transformed into a pre-defined data model, with healthy regularity, strict format, poor scalability, and low flexibility. Semi-structured data can be considered as an extension of structured data, which has better flexibility and extensibility. Therefore, we will treat semi-structured and structured data as the same. 

To overcome the drawbacks of questionnaire surveys, such as small survey scope, hard to update, time-consuming, and labor-intensive, Li et al. \cite{RN77} proposed a machine learning-based affective design dynamic mapping approach (MLADM). Firstly, collect Kansei words from literature and selected them by manual clustering. Secondly, get product features and pictures from shopping websites. Then, automatically generated online questionnaires. Finally, four machine learning algorithms were used to construct the relationship model. Although the MLADM method can predict product feeling, it still relies heavily on the expert. Besides, online questionnaire data also possess highly subjective. Some studies explore design knowledge from objective data instead of questionnaires. For instance, Jiao et al. \cite{RN78} established a database to record affective needs, design elements, and Kansei words from past sales records and previous product specifications. They adapted association rule mining to build the relationship model and take it as the product design inference pattern.

Limited by data sources' openness, it is not easy for designers to obtain structured data, which hinders product design research. Compared with structured data, unstructured data have advantages in volume and openness. Nearly 95\% of big data are unstructured \cite{RN79}. Although unstructured data is amorphous and complicated, it still attracts scholars' attention due to its implicit information and value. In Section 3.3-3.6, we will respectively introduce the application of textual data, image data, audio data, and video data in product design.

 \subsection{Product design based on textual data}

\begin{figure}
  \centering
  \includegraphics[width=15cm]{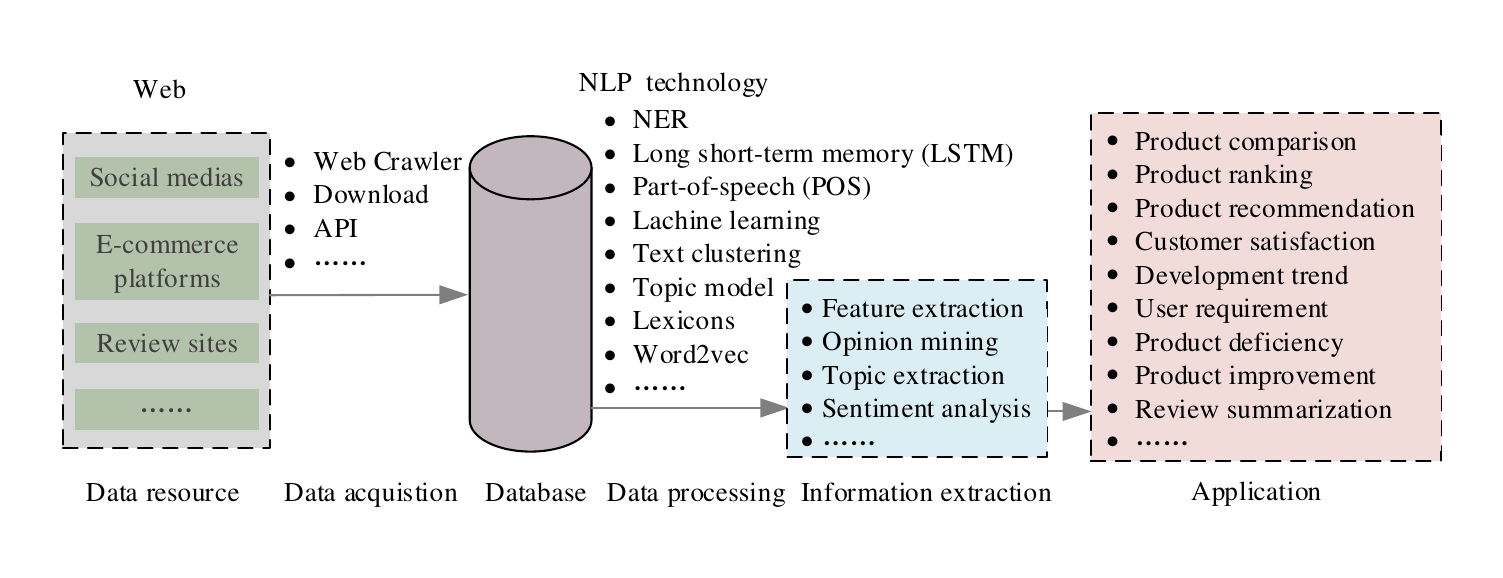}
  \caption{The typical processing flow of textual data. It includes five parts, namely data acquisition, data processing, information extraction, and application.}
  \label{fig:fig7}
\end{figure}

The text exists in many phases of the product lifecycle. For product design, the most valuable is the user experience feedback after purchasing the product. More and more users have posted their opinions of products, services, and events on Twitter, Facebook, microblogs, blogs, and e-commerce websites in recent years \cite{RN129, RN128, RN130, RN127, RN131, RN123, RN95}. Online reviews of e-commerce websites are given by consumers who purchased them. So the reviews have  high reliability and authenticity \cite{RN132}. According to research: 90 \% of customers take online reviews as a reference, 81\% indicate it is helpful for purchase decisions, and even 45\% consulted online reviews in physical stores \cite{RN85, RN118}. The online review has become an important influence factor in consumers' behaviors and a reference for enterprises to improve their products \cite{RN114, RN105}. It helps to understand customer satisfaction \cite{RN109, RN104, RN133, RN103, RN96}, grasp user requirements \cite{RN108, RN93, RN86}, find product deficiencies \cite{RN82, RN119, RN125}, propose improvement strategies \cite{RN262}, compare competitive products \cite{RN94}, product recommendation \cite{RN89, RN111, RN110, RN113, RN87, RN84}, etc. For instance, in order to identify innovation sentences from online reviews, Zhang et al. \cite{RN106} proposed a deep learning-based approach. Jin et al. \cite{RN94, RN126, RN114, RN122} devoted to filtering out helpful reviews. Wang et al. \cite{RN120} presented a heuristic deep learning method to extract opinions and classified them into seven pairs of affective attributes (namely, like-dislike, aesthetic-inaesthetic, soft-hard, small-big, useful-useless, reliable-unreliable, and recommended-not recommended). Kumar et al. \cite{RN92} combined reviews and electroencephalogram (EEG) signals for product rating prediction. Xiao et al. \cite{RN85} proposed a marginal effect-based KANO model (MEKM) to categorize customer requirements. Simon et al. \cite{RN112} explored product customization based on online reviews. We summarized the typical processing flow of textual data in Figure \ref{fig:fig7}.

The rapid development of NLP is the key to product design based on textual data, which including topic extraction, opinion mining, text classification, sentiment analysis, and text clustering \cite{RN99}. 
 
  \textbf{\emph{Product attributes extraction}} In the field of product design, topic extraction can be used to extract product attributes (e.g., the screen, battery, weight, and camera of a smartphone) and opinions. Each product has multiple attributes, and consumers have different preferences and evaluations for each attribute of each product \cite{RN111}. Therefore, it is necessary to extract product attributes from online reviews. Product attributes are usually nouns or noun phrases in online review sentences \cite{RN100}. Most studies adapt the part-of-speech (POS) tag to generate tags (whether the word is a noun, adjective, adverb, etc.) and taking all nouns as attribute candidates. Then prune candidates according to term-frequency (TF) \cite{RN122, RN90}, term frequency-inverse document frequency (TF-IDF) \cite{RN124, RN88}, dictionary \cite{RN134}, manual definition \cite{RN133}, clustering \cite{RN81, RN105}, etc., to obtain product attributes. Moreover, since product attribute is sensitive to the domain, some studies regard it as a domain-specific entity recognition problem. For instance, Putthividhya and Hu \cite{RN83} used named entity recognition (NER) to extract product attributes.
  
 The methods mentioned above can only extract explicit attributes and do not work for implicit attributes. Whether it is explicitly mentioned is the main difference between explicit and implicit attributes. For example, "Laptop battery is durable, the laptop is expensive." The "battery" is explicitly mentioned as an explicit attribute, while "expensive" is related to price and is an implicit attribute. The implicit attribute extraction methods can be divided into unsupervised, semi-supervised, and supervised \cite{RN97}. Xu et al. \cite{RN98} used an implicit topic model, which incorporated pre-existing knowledge to select training attributes,to build an implicit attribute classifier based on SVM. Due to the lack of a training corpus, they annotated a large number of online reviews. Kang et al. \cite{RN117} proposed the unsupervised rule-based method, which can extract subjective and objective features from customer reviews.

 \textbf{\emph{Opinion mining}} In product design, opinion mining aims to extract descriptive words that express subjective opinions, also called opinion words. The method of extracting opinion words can be divided into two categories. The first one is based on the co-occurrence of attributes and opinion words, taking the adjectives adjacent to product attributes as opinion words. The other is syntactic analysis, which is based on the structure and dependency of the review sentences. Hu and Liu \cite{RN102} divided product attributes into frequent and infrequent. The frequent attributes were obtained by POS and association mining and taking nearby adjectives as opinion words. The infrequent attributes were obtained through the reverse search of opinion words. 
 
Since different reviewers may have different expressions of the same product attributes, Jin et al. \cite{RN94} provided a group of manually defined synonyms for clustering to improve WordNet. WordNet is an English lexical database, which organizes the words in terms of synonyms and antonyms. In our previous research \cite{RN261}, we clustered similar opinion words based on word2vec. According to the grammatical rules, Wang et al. \cite{RN91} restructured raw sentences to extract attributes-opinions from online reviews. However, different languages have different sentence structure rules, so it is hard to expand to other languages.

In addition to the two types of opinion word extraction, the topic model also shows excellent potential. Bi et al. \cite{RN109} used latent Dirichlet allocation (LDA) to extract customer satisfaction dimensions from online reviews and proposed the effect-based KANO model (EKM). LDA is a topic model that can group synonyms into the same topic and obtain its probability distribution. Wang et al. \cite{RN101} applied the long short-term memory (LSTM) model to extract opinion words from raw online reviews and mapped customer opinions to design parameters through deep learning. However, the topic model ignores the fine-grained aspects.

 \textbf{\emph{Sentiment analysis}} In product design, sentiment analysis is mainly to identify customers' affective attitudes towards the object product and labeled them as positive, negative, or neutral. According to the different scopes, sentiment analysis can be performed in three levels: (i) document-level; (ii) sentence-level; and (iii) aspect (attribute)-level \cite{RN249, RN248}. The document-level is for all reviews in a document, and it is coarse-grained. The sentence-level is for a sentence, and it is between coarse-grained and fine-grained. The aspect-level is for a certain attribute of a product, and it is fine-grained. 
 
The method of sentiment analysis can be divided into three categories: (i) machine learning method; (ii) lexicon-based method; and (iii) hybrid method \cite{RN250, RN92, RN251, RN120}. Machine learning methods regard sentiment analysis as a classification problem, taking sentiment polarities as labels and adapt Recurrent Neural Network (RNN) \cite{RN106}, support vector machines \cite{RN253, RN252}, conditional random field (CRF) \cite{RN249}, Neural Network \cite{RN248}, etc., to construct classifiers with text features. Lexicon-based methods identify sentiment orientations by referring to pre-defined lexicons such as LIWC, HowNet, and WordNet \cite{RN254, RN252}. These lexicons contain sentiment-related terms and their polarity. Lexicon-based methods relying on the quality of the sentiment lexicon, so some scholars are committed to improving the coverage, domain adaptation, and continuous updating of lexicons. For instance, Cho et al. \cite{RN255} constructed a comprehensive lexicon by merging ten lexicons, such as WordNet, General Inquirer, and AFINN. Araque et al. \cite{RN256} proposed a sentiment classification model that used the semantic similarity measure combined with embedding representations. They identify sentiment orientations by computing the semantic distance between the input word and lexicon word rather than keyword matching. Marquez et al. \cite{RN258} analyzed several lexicons and how they complement each other.

The collection of the lexicon can be divided into three categories: (i) manual, (ii) dictionary-based, and (iii) corpus-based. The manual method is labor-intensive and time-consuming; the dictionary-based method taking a set of opinion words as seeds to search in existing dictionaries; the corpus-based method expands lexicons based on the information in the corpus, and the expanded lexicon can be domain-specific \cite{RN259}. Machine learning methods tend to be more accurate, lexicon-based methods are more general, and hybrid methods combine them. For instance, Dang et al. \cite{RN260} proposed a lexicon-enhanced method for sentiment classification combining machine learning and semantic-orientation approaches. Their results showed that the hybrid method could improve sentiment classification performance significantly. 
 
By reviewing literature relate to product design based on textual data, we found it had been an important research topic addressed by many scholars in recent years. NLP technology is used to extract product attributes and sentiment features from the text, and apply them to understand user requirements, evaluate products, and grasp product development trends. Requirements and evaluates are the core and foundation, but this does not mean that the rest design process is not important. Most existing research ignores this point, making it more like an improved design than an original innovation design. In other words, product design based on text still has a vast space for development. Undoubtedly, for product design, the emergence of textual data is of great significance. Compared to traditional methods, textual data provides rich information with higher quality and require less time to acquire and process.

\subsection{Product design based on image data}
\begin{figure}
  \centering
  \includegraphics[width=15cm]{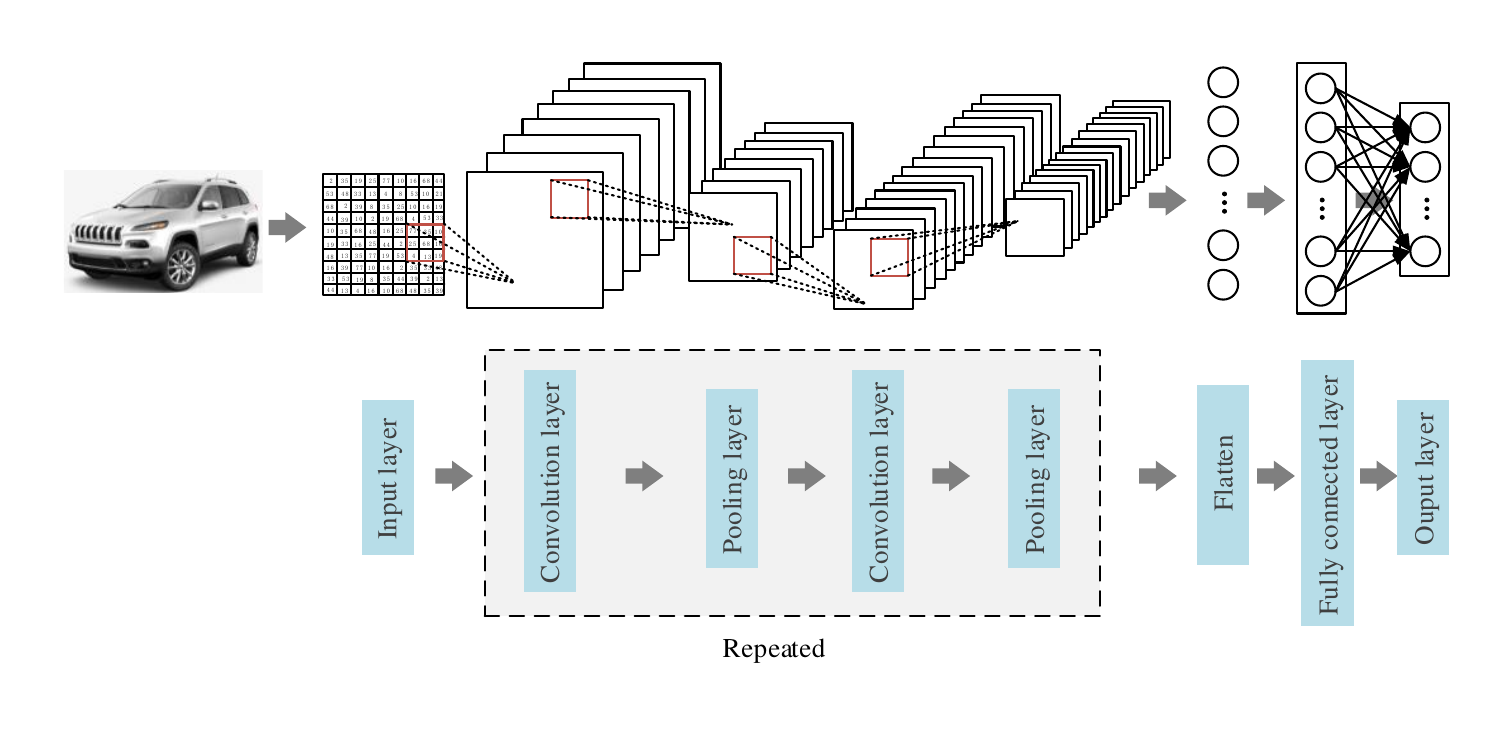}
  \caption{The structure of convolutional neural networks (CNN). CNN consists of the input layer, convolution layer, pooling layer, fully connected layer, and output layer. The input image will be converted into the pixel matrix in the input layer. The convolution layer involves some filters, and different filters get triggered by different features (e.g., semicircle, triangle, quadrilateral, red, and green) in the input image. Each filter will output a feature map, and the pooling layer will reduce its dimensionality. The fully connected layer and the output layer are the same as the artificial neural network.}
  \label{fig:fig8}
\end{figure}

Product images are intuitive expressions of products that contain color, texture, and shape information, which are of great significance to design. The information provided by images is easy to grasp by our eyes, which is a significant advantage. What sees is what gets. Image data's continuous growth forms big image data, which has become another vital information carrier after textual data. In recent years, CNN has made breakthroughs in image recognition, classification, and segmentation \cite{RN203, RN267, RN213}. Its robust feature learning ability has attracted attention in various fields. Figure \ref{fig:fig8} shows the structure of CNN. We divide the role of image data in product design into two types: inspiration and generation. The former inspires designers by existing product images, while the latter can directly generate a new product image based on large amounts of images.

  \textbf{\emph{Stimulating the creative inspiration of designers.}} Existing product images are helpful to inspire designers to come out with new ideas and initial design schemes more quickly \cite{RN137}. The image retrieval, match, and recommendation are key for visually inspire \cite{RN167, RN150, RN168, RN178}. Designers and users put forward their requirements through images and text, search for related product images from databases or e-commerce websites, and the matched images will be recommended to designers and users as design references. The retrieval input can be text, images, or both of them \cite{RN186, RN207, RN164, RN191, RN159}. For product, the input image provided by designers and users may be taken by their phone on the street or in a store, which is quite different from image databases and e-commerce websites in terms of shooting angle, condition, background, or posture \cite{RN165, RN149, RN188, RN158, RN202}. Therefore, product retrieval is more complicated than a simple image retrieval \cite{RN155, RN157, RN195}. To search clothing images taken in the street on e-commerce websites, Liu et al. \cite{RN201} adopted the human detector to located 30 human parts, then considered a mapping with a sparsely coded transfer matrix to ensure the difference between these two distributions will not affect the retrieval quality. Compared with photos taken on the street, free-hand sketches are more abstract. Yu et al. \cite{RN199} introduced two new datasets with dense annotation and built a deep network learned with triplet annotations to achieve retrieval across the sketch/image gap. Ullah et al. \cite{RN211} used a 16-layer CNN model (VGG16) to extract product features from images and employed Euclidean distance to measure their similarities. Ristoski et al. \cite{RN134} used CNN to extract image features and match them with text embedding features to improve product matching and categorization performance. Liang et al. \cite{RN147} proposed a joint image segmentation and labeling framework to retrieve clothing. Specifically, they grouped superpixels into regions and selected confident foreground regions to train the E-SVM classifier, and propagated segmentations by applying the E-SVM template overall images. 
 
In addition to serving as a reference, image data can also help track fashion trends \cite{RN190}, identify user preferences \cite{RN177, RN209}, product matching recommendations \cite{RN171, RN156, RN169, RN185}, and prevent infringement \cite{RN137}. For instance, Hu et al. \cite{RN210} established a furniture visual classification model containing 16 styles (e.g., Gothic style, Modernist style, and Rococo style), which combined image features extracted by CNN with handcrafted features. This model can help users understand their preferred styles. Most products do not exist alone, and their compatibility needs to be considered during the design process. Therefore, designers and users also need to be recommended with a set of compatible products. Aggarwal et al. \cite{RN183} used Siamese networks to assess the style compatibility between pairs of furniture images. They also proposed a joint visual-text embedding model for the recommendation, derived from furniture type, color, and material. Due to the narrow application of pairwise compatibility, Laura et al. \cite{RN181} built a graph neural network (GNN) model to capture multiple items' interactions. Their GNN model comprised a deep CNN to extract image features. In addition to evaluating the compatibility of multiple furniture in style, color, material, and overall appearance, they also applied the GNN model to solve the fill-in-the-blank task. For example, according to a given desk, cabinet, chair, and mirror, recommend the most matching bed from multiple alternatives.

\begin{figure}
  \centering
  \includegraphics[width=15cm]{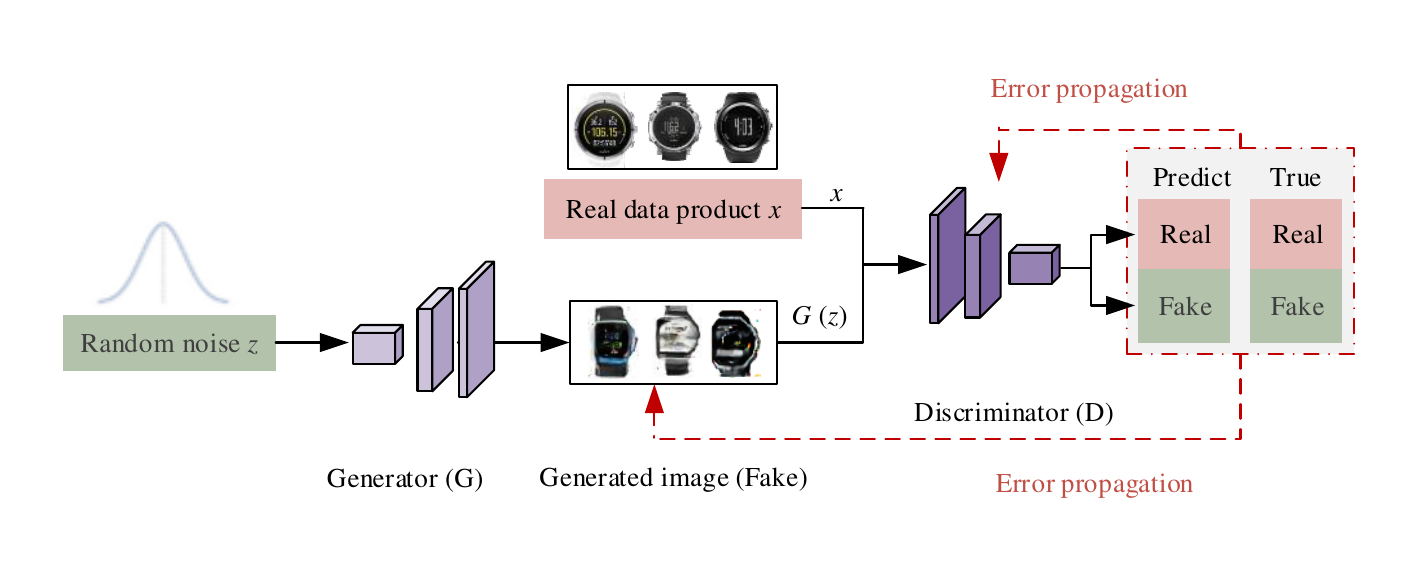}
  \caption{The structure of GAN. Setting $x$ as the real product and $z$ as the random noise, $G(z)$ is the synthetic data generated by the generator $G$. By the way, images can be regarded as a kind of noise distribution. Both $x$ and $G(z)$ are inputted to the discriminator $D$ to predict the data is real or fake. If the predicted result is correct, the error will be transferred to $G$ for improvement; otherwise, it will transfer to $D$ for improvement. Eventually, $G$ could capture the statistical distribution of $x$, and $G(z)$ can deceive $D$. $G(z)$ is the generated design scheme that contains product features and is different from the real product.}
  \label{fig:fig9}
\end{figure}

  \textbf{\emph{Generating new product images based on image data.}}  By reviewing a large amount of literature on generative product design, James and Jason \cite{2020Progress} found that deep learning techniques can facilitate product design. Scholars have made some explorations in applying generative models to product design. As representative generative models, both GAN and style transfer take images as input and output. GAN is an unsupervised model composed of two neural networks: the generator and discriminator \cite{RN247}. Both generator and discriminator are trained simultaneously and are improved in their competition. Figure \ref{fig:fig9} shows the structure of GAN. We summarized the contribution of GAN to product design as schemes generation \cite{RN212, RN145, RN151}, text-to-image synthesis \cite{RN198}, generative transformation \cite{RN184}, collocation generation \cite{RN170, RN197, RN189}, sketch acquisition \cite{RN192, RN246}, colorization \cite{RN136, RN162, RN206}, and virtual display \cite{RN140, RN193}.
  
The automatic generation of product schemes is based on a large number of existing product images. GAN can automatically learn the distribution of product features from images, and the discriminator gradually optimizes the generator to generate product schemes. For instance, Li et al. \cite{RN137} used GAN to generate new design schemes for smartwatches. They built smartwatch data with 5459 images and used it for GAN training. The trained GAN can automatically capture feature distribution and generate new smartwatch images that are not in the training set. Furthermore, they also compared the results of GAN and its extension models, including deep convolution GAN (DCGAN), least-squares GAN (LSGAN), and Wasserstein GAN (WGAN). To generate collocation clothing images, Liu et al. \cite{RN205} proposed Attribute-GAN, which involves a generator and two discriminators. GAN only provides one view of the generated scheme, which is insufficient for design. To synthesize an arbitrary view of product images, Chen et al. \cite{RN140} proposed conditional variational GAN (CVGAN). CVGAN is very important for virtual display, especially for designs with high requirements of shape and appearance. For example, clothing design, which needs to show the effect of try-on. In addition to generating product images, GAN and its extended models can also apply to layout generation, such as indoors \cite{RN172, RN174} and web pages \cite{RN144}.

Text-to-image synthesis is another useful technique for product design, but existing studies are rare. Kenan et al. \cite{RN198} presented an enhanced attentional GAN (e-AttnGAN) for synthesizing fashion images from text descriptions (e.g., long sleeve shirt with red check pattern). Most studies focus on improving synthesis technology \cite{RN153, RN173, RN180, RN143}, and text-to-image has not attracted designers' attention yet. We believe that text-to-image will become an important development direction for product design in the future. Users just need to express their requirements in texts, and their desired product images will be automatically generated.

Generative transformation and colorization can be regarded as the image-to-image problem, where the image is input, and the modified image will be output \cite{RN187, RN166, RN146}. The generative transformation aims to convert one product image into another one. During the transformation, a sequence of intermediate images will be automatically generated, which can be used as new design schemes or to inspire designers. Zhu et al. \cite{RN179, RN175, RN208} developed a product design assistance system based on GAN and showed three applications: (i) change product shape and color by manipulating an underlying generative model. (ii) Generate new images from users' scribbles. (iii) Generative transformation of one product picture to another product. For example, taking a short black boot and a long brown boot as inputs, the former can be automatically converted into the latter, and intermediate shoe images are shown. Nikolay et al. \cite{RN204} proposed the conditional analogy GAN (CAGAN) to automatically swap clothes on models. The sketch already contains the structural and functional features of the product. Once colored, a preliminary design scheme is completed. Compared with other features, color has an intuitive influence, and users have different preferences for it. GAN can automatically color the sketch and quickly generated multiple alternatives. Liu et al. \cite{RN139} established an end-to-end GAN model for ethnic costume sketches colorization. Their results showed that GAN has an excellent ability to learn the color rules of ethnic costumes. Sreedhar et al. \cite{RN151} proposed a car design system based on GAN that supports single-color, dual-color, and multiple-color coloring from a single sketch.

GAN and its extended models have brought significant changes to product design, but the generated images are blurry, lacking details, and low quality. To solve the problem of losing detail and unnatural display, Lang et al. \cite{RN197} proposed Design-GAN, which introduced the texture similarity constraint mechanism. Oh et al. \cite{RN161} combined GAN with topology optimization to guarantee the quality of two-dimensional wheel images. They have restricted wheel image as non-design domain, pre-defined domain, and design domain for topology optimization. Their results show that the optimum is far from the initial design.

\begin{figure}
  \centering
  \includegraphics[width=15cm]{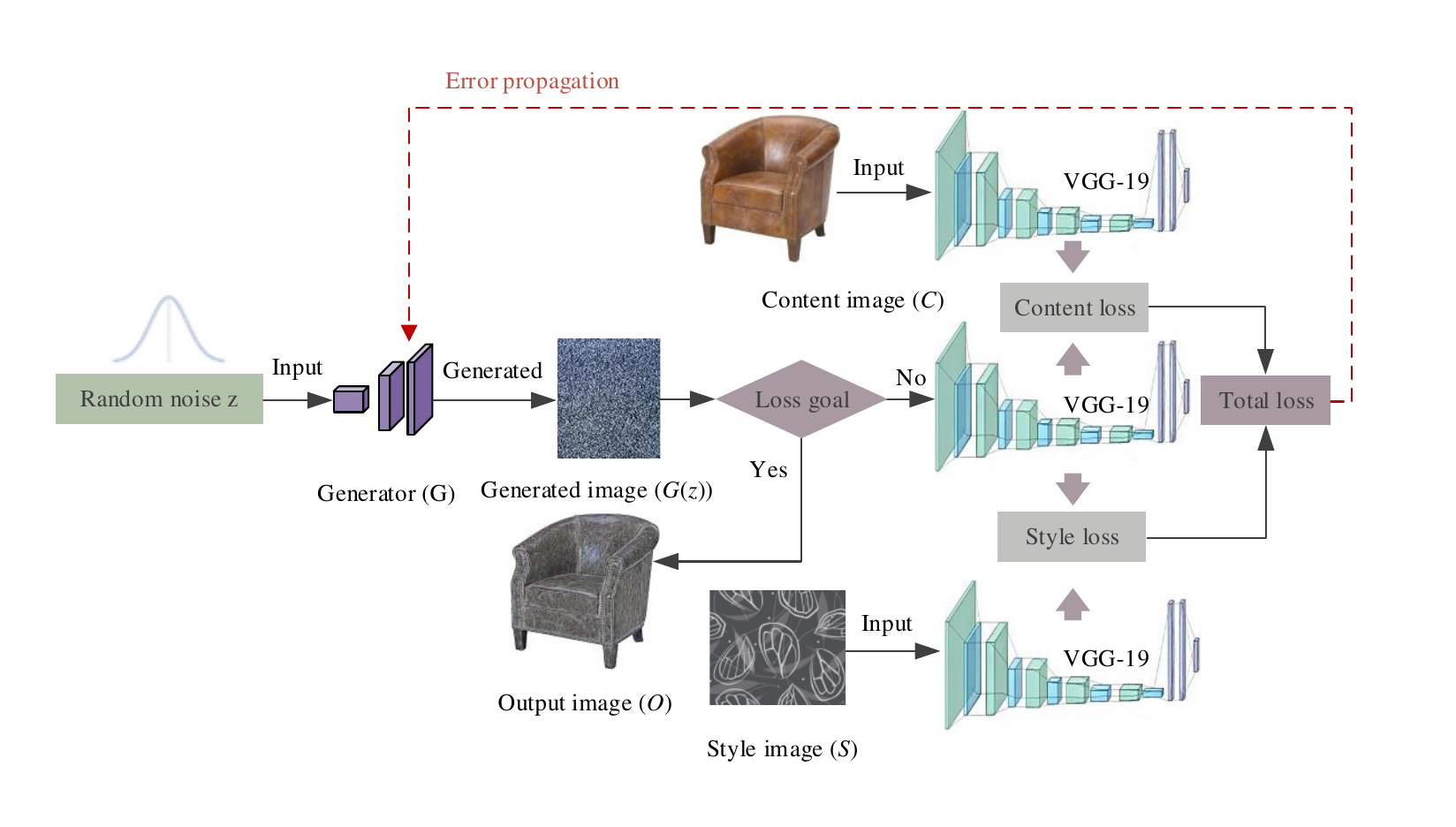}
  \caption{Neural style transfer. Setting z as the random noise, G (z) is the synthetic image generated by the generator (G). The pre-trained VGG 19 is used to calculate the style loss between G (z) and the style image (S), the content loss between G (z) and the content image (C). Minimize the total loss that consists of style and content loss to optimize G. Once the loss goal is reached, G (z) will be output and marked as O. O can preserve the content feature of C and the style feature of S.}
  \label{fig:fig10}
\end{figure}

Neural style transfer is also a deep generative model that realizes image generation by separating and reconstructing images' content and style. Nevertheless, the neural style transfer does not mean merely overlapping a content image and a style image. Its implementation is related to the features learned by CNN. Figure \ref{fig:fig10} shows the idea of neural style transfer.

In 2015, Gatys et al. found that the image style and content were separable in CNN, and they can be manipulated independently. Based on this finding, Gatys et al. \cite{RN135} proposed a neural style transfer algorithm to transfer the style of famous artworks. Huang and Serge \cite{RN141} proposed a novel adaptive instance normalization (AdaIN) layer, which aligns the mean and variance of content features with style features to realize arbitrary style transfer. Inspired by their method, in our previous research \cite{RN194}, we attempted to apply neural style transfer to product design scheme generation. We combined Kansei Engineering with neural style transfer, and the result showed that the semantic of the generated product had been enhanced. Later, we developed a design system that allowed users to input arbitrary images and output schemes in real-time \cite{RN245}. Wu et al. \cite{RN148} combined GAN with neural style transfer to generate fashionable Dunhuang clothes. GAN is used to generate clothes shapes based on the open-source Fashion Mnist dataset, and neural style transfer is used to add Dunhuang elements. Neural style transfer has only developed for five years, and there are not many applications in product design. While existing researches are taking the entire image as content, which limits the flexibility of product design. It is possible to transfer different styles to different parts of a product through masks \cite{RN141}. For example, for a coat, three style images can be transferred for the collar, pocket, and sleeve, respectively. Neural style transfer still has great application potential in product design.

By reviewing the related research of image data, we found that it opened up new directions for the product design field, especially in schema generation and inspiration. GAN and its extended models are unstable for design scheme generation, with low image quality and strong randomness. However, Sohn et al. \cite{RN142} demonstrated that artificial intelligence and GAN enhanced customer satisfaction and freshness. Their results showed the potential of GAN in product design. Neural style transfer is also used to generate product schemes with good image quality, strong interpretability, and strong controllability. Compared to GAN and its extended models, neural style transfer has a weaker innovation ability, and it can be regarded as a combination design of product shape, texture, and color. Generation promises to be a cost-effective way to product designs and support designers to create design schemes quickly.

In general, image data have broken old ideologies of traditional product design. The literature regarding product design based on generative models is still relatively sparse. Moreover, most of them only focus on technical aspects, lacking product design knowledge \cite{RN182, RN244}. It is not feasible to just replace different training image data to realize design \cite{RN160, RN196}. Product design has its particularities and demands. Traditional product design methods and design knowledge cannot be abandoned, and they should be combined with big data. We think that is maybe where product design is headed.

\subsection{Product design based on audio data}

Big audio data \cite{RN228} is information with sound or voice. The customer center brings together users' complaints, inquiries, and suggestions during the product service cycle \cite{RN243}. Audio feedback is useful for user requirements acquisition, recommendation, product evaluations, product design, and improved design. Since the literature regarding product design based on big audio data is still almost blank, we will introduce several key technologies that may be used in the product design field in the future, such as speech recognition, speaker identification, and emotion recognition.

\begin{figure}
  \centering
  \includegraphics[width=15cm]{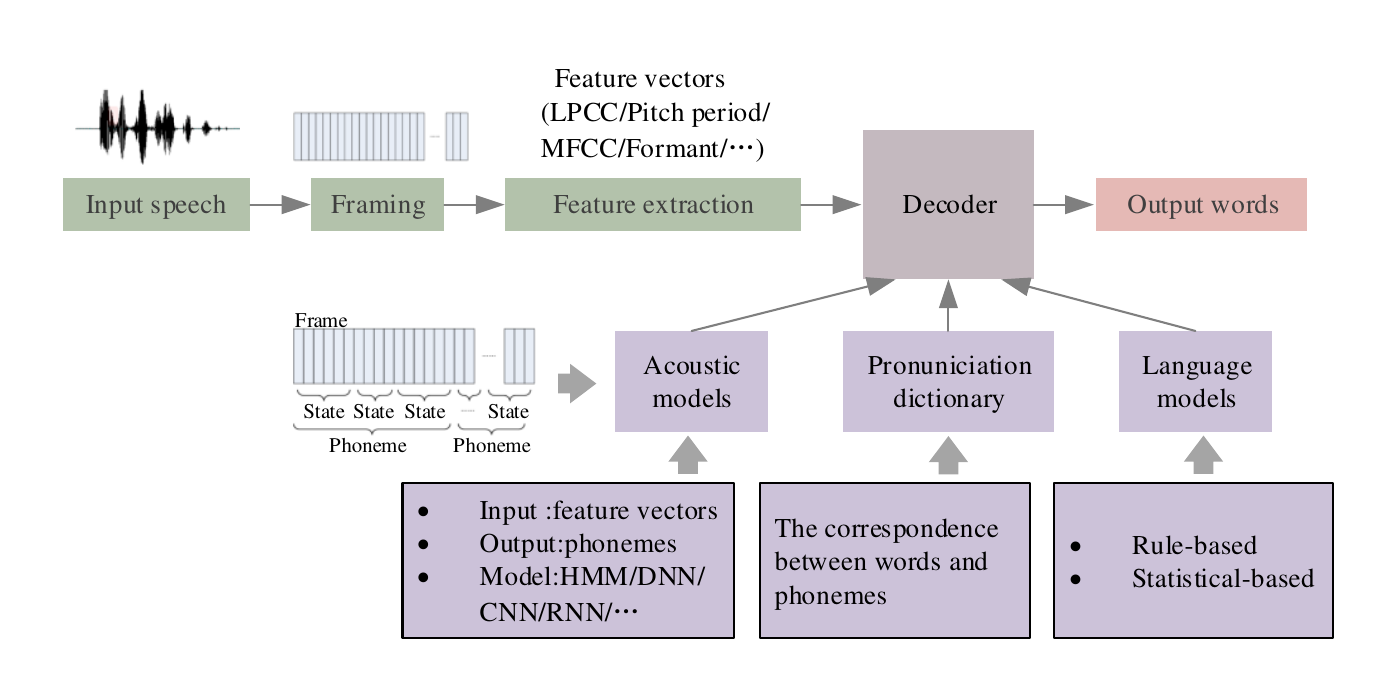}
  \caption{The process of speech recognition. Framing the input speech into many frames and transfer the waveform to extract feature vectors. The acoustic model is then used to convert features into phonemes, matched to words through the pronunciation dictionary. Finally, it eliminates the confusion of homophones by the language model.}
  \label{fig:fig11}
\end{figure}

Some studies manually record telephone complaints as text to avoid direct processing of audio signals \cite{RN222, RN216}. However, manual recording heavily relies on the recorder, and it is a labor-intensive and time-consuming process. In comparison, speech recognition can accomplish this task automatically \cite{RN223}. Speech recognition, also known as automatic speech recognition (ASR), computer speech recognition, and speech-to-text, is converting human speech into computer-readable information, which usually refers to text, but it may also be binary codes or character sequences \cite{RN221}. The process of speech recognition is shown in Figure \ref{fig:fig11}. Speech recognition widely used in mobile communication, search engines, human-computer interaction, etc. \cite{RN220}

\begin{figure}
  \centering
  \includegraphics[width=15cm]{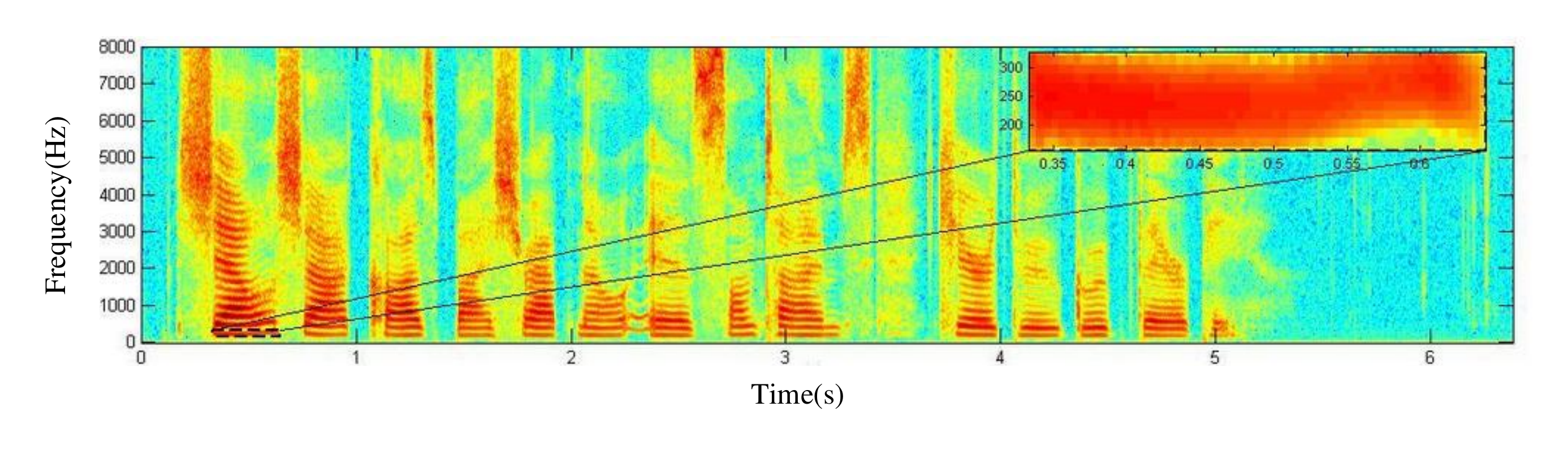}
  \caption{Spectrogram. The horizontal axis represents time, and the vertical axis represents frequency. The amplitude of speech at each frequency point is distinguished by color. }
  \label{fig:fig12}
\end{figure}

Speaker identification, also known as voiceprint recognition, is a technology to identify a speaker by their speech. Each voice has unique characteristics determined by two factors, the size of the sound cavity and how the vocal organs are manipulated. Like fingerprints, voiceprints also need to be collected in a database in advance. The spectrogram is drawn by the amplitude of the short-time Fourier transform (STFT) of the audio signal (Figure \ref{fig:fig12}) \cite{RN219}. Voiceprint recognition can be achieved by extracting speaker parameters (such as pitch frequency and formant) and using machine learning methods. Voiceprint recognition has been widely applied in biometric authentication, crime forensics, mobile payment, social security, etc. However, factors such as the variability of the voiceprint, the different quality of audio acquisition equipment, and environmental noise interference make the voiceprint recognition tasks highly challenging.

As a significant audio processing and emotion computing branch, speech emotion recognition aims to identify the speaker's emotions \cite{RN225}, such as anger, sadness, surprise, pleasure, and panic. Speech emotion recognition can be regarded as a classification problem, and the selection of emotion feature is the core \cite{RN217, RN218}. We summarized commonly used acoustic parameters in Table \ref{tab:table2}. Besides, the spectrogram is the audio signal's image expression, so CNN can be used to automatically learn the features to perform speech emotion recognition \cite{RN224}.

\begin{table}
 \caption{Acoustic parameters.}
  \centering
  \begin{tabular}{ p{5cm} p{8cm}}
    \toprule
    Category&Parameter\\
    \midrule
    Prosody parameter&Duration\\
 &Pitch\\
 &Energy\\
 &Intensity\\
    Spectral parameter&Linear predictor coefficient (LPC)\\
 &One-sided autocorrelation linear predictor coefficient(OSALPC)\\
 &Log-frequency power coefficient(LFPC)\\
 &Linear predictor cepstral coefficient(LPCC)\\
 &Cepstral-based OSALPC(OSALPCC)\\
 &Mel-frequency cepstral coefficient(MFCC)\\
    Sound quality parameter&Format frequency and bandwidth\\
 &Jitter and shimmer\\
 &Glottal parameter\\
    \bottomrule
  \end{tabular}
  \label{tab:table2}
\end{table}

With the popularity of artificial intelligence products, more and more attention has been paid to audio signal processing technology, especially speech recognition and emotion recognition, which directly affect the interactive experience and are the key to customer choice. Although audio data has great potential in product design, due to the limitations of processing technology and data quality, there is still a long distance between its theory and practical applications.

 \subsection{Product design based on video data}
Videos of product usage are also valuable information source for product design, which can be viewed repeatedly without restriction, making it easier to obtain hard-to-find but essential information. For instance, by watching the cooking videos of homemakers, Nagamechi found that standing up to take food made them feel more comfortable than bending over. Moreover, the refrigerator compartment is used more frequently than the freezer compartment. Therefore, Nagamechi improved the traditional refrigerator structure, changing the upper layer to a refrigerator compartment and the lower layer to a freezer compartment \cite{RN242}. This structure is still in use today. Although video data is useful, manual viewing is time-consuming and only suited for small data volume scenes. In comparison, intelligent video analysis can realize motion detection, video summarization, video retrieval, color detection, object detection, emotion recognition, etc., \cite{RN231, RN227, RN228, RN232} independent of human engagements. Intelligent video analysis makes it possible to apply big video data in user requirement acquisition, user behavior observation, experience improvement, and product virtual display. Intelligent video analysis establishes a mapping relationship between images and image descriptions so that computers can understand video through digital image analysis.

 \textbf{\emph{Product detection.}} Compared with a single image, object detection in the video has a temporal context, which helps to solve the redundancy between consecutive frames, motion blur, video unfocused, partial occlusion, singular postures, etc. Li et al. \cite{RN236} proposed a novel method to annotate products in videos. They identified keyframes and extracted SIFT features to generate BOVW histograms. Then compared them with products' visual signature for annotation. Zhang et al. \cite{RN229} proposed a framework to identify clothes worn by stars in videos. They adopted DCNN to detect related images from frames, including human body detection, human posture selection, human pose estimation, face verification, and clothing detection. Moreover, Zhang et al. \cite{RN239} linked clothes worn by stars with online shops for clothing recommendations. To improve the matching results, Chen et al. \cite{RN241} used image feature network (IFN) and video feature network (VFN) to generate deep visual features for shopping images and clothing trajectories in videos.

 \textbf{\emph{User behavior observation.}} Understanding user behavior is essential for product design and improving user experience. To explore user behaviors in VR spherical video streaming, Wu et al. \cite{RN226} recorded a head tracking dataset. Babak et al. \cite{RN240} assessed the usability of products from videos. The core idea is to perform temporal segmentation of video containing human–product interaction and automatically identified time segments in which humans encounter difficulties. They took water faucet design as an example and used optical flow for motion detection. Optical flow, temporal difference, and background subtraction are commonly used for motion detection in videos \cite{RN230}.
 
 \textbf{\emph{Product virtual display.}} Compared with images, the video-based virtual display has higher demands, such as spatiotemporal consistency. There is no doubt that the video-based virtual display can provide a more comprehensive understanding for customers. Liu et al. \cite{RN234} presented a parsing model to predict human poses in the video. Dong et al. \cite{RN235} proposed a flow-navigated warping GAN (FW-GAN) to generate a try-on video conditioned on a person image, the desired clothes image, and a series of target poses. The interactive display will promote product understanding and evoke enjoyment during the experience \cite{RN233}. Therefore, An et al. \cite{RN237} designed a video composition system that displays products on mobile phones in an interactive and 3D-like way. It can automatically perform rotation direction estimation, video object segmentation, motion adjustment, and color adjustment.

 Current intelligent video analysis technology can only analyze some simple activities, while complex activities still have difficulties. Structuralization is a vital obstacle to video analysis, which identifies features in the video through background modeling, video segmentation, and target tracking operation. In addition, complete video data acquisition requires the cooperation of multiple cameras. It is difficult to get consistent installation conditions for each camera and perform a continuous and consistent visual analysis of moving targets in multiple videos. Moreover, the motion detection technology is not mature. Existing motion detection can only achieve target entry detection, departure detection, appearance detection, disappearance detection, wandering detection, trailing detection, etc. The product use process is complicated, and the information extracted at present is of little help to the product design. In summary, audio data has great potential in product design, but its development and application are restricted due to processing technology and data quality.

 \subsection{Summary of opportunities}
 Big data has big potential in data driven product design. There are two types of data. Structured data have the advantage of strong pertinence and high-value density, which, however, have limitations in openness and volume. It is difficult for individuals to obtain large-scale structured data. Even enterprises require lots of workforce and resources, involving manual collection, sorting, summarizing, and storage in standard databases. On the other hand, unstructured data enjoy the high openness and volume compared to structured data. Besides, it has great advantages in terms of veracity, velocity, value, and variety, all may strongly contribute to data driven product design.

 We have detailed the common data types in the product lifecycle, including text, image, audio, and video. Textual data and image data are widely used in product design due to easy data acquisition, mature processing technology, and high data quality. Specifically, textual data can overcome the shortcomings of traditional survey methods (i. e., a small amount of data, small survey scope, hard to update, time-consuming, and labor-intensive). To a certain extent, it guarantees the authenticity, reliability, and timeliness of the data. Image data are also useful for the survey, and it is more prominent in design schemes display and visual inspiration. The color, texture, sharp information in images can inspire designers to obtain preliminary schemes. Moreover, generative models can generate new product designs from image data with varying colors, structures, textures, etc. Automatically generating design images is revolutionary in product design, which has never been realized in traditional product design methods. 
 
 Video data is also visible type of information, which is highly valuable in user behavior observation and product virtual display. Compared with images, video can provide customers with a comprehensive understanding and allowing them to interact with products virtually. While it is unfortunate that both video data and audio data are still in their infancy and there are many difficulties to be solved (especially in terms of processing technology and data quality) before they can be applied in product design, their potential is huge. Although more and more advanced technologies are developed for audio and video processing, few researches have employed them in product design. However, these limited studies warrant a discussion to motivate further research in the field.
 
 Big data has created a host of benefits to product design, but existing research still has some shortcomings which call for further research. There are two research directions that are especially promising: (i) Fusion of different types of data. Existing research is confined to only one type of data (i.e., text, image, video, audio, and location). These data co-exist in different stages of the product lifecycle. Furthermore, multi-modal data fusion will increase the persuasiveness and the effect of the extracted information. (ii) Synergistic exploitation of big data techniques and design domain knowledge. Previous efforts have focused on technological breakthroughs, which are critical, especially for data that lack mature processing technologies, such as audio and video. Nevertheless, this does not mean that domain knowledge in the field of product design should be ignored. The knowledge accumulated in traditional product design is precious and helpful. Big data and traditional product design methods are not contradictory but cooperative and complementary. Only by combining them can we better capture user requirements and improve the success rate of design.
 
 \section{Conclusions}
 \label{sec:Conclusions}
 In the era of the knowledge-driven economy, customer demands are more diversified and personalized and the lifecycle of products are becoming shorter, especially in the usage phase. Therefore, product innovation requires the assistance of interdisciplinary knowledge, the support of powerful techniques, and the guidance of innovation theories that break conventional wisdom. 
 
 At present, big data is one of the most potential resources for promoting innovations for the whole product lifecycle. Here we have presented a comprehensive review of existing big data driven production design studies to facilitate researchers and practitioners comprehending the latest development of relevant studies and applications. Firstly, we introduced several representative traditional product design methods and revealed their shortcomings.  Secondly, we illustrated big data's potential in solving the challenges in modern product design, especially in terms of user requirements acquisition, product evaluation, and visual display. The current and potential application of textual data, image data, audio data, and video data in product design have been reviewed in detail. Since audio data is still in its fledgeless stage, we focused on possible processing technologies and workflow that may be applied in future. Finally, the deficiencies of existing studies and future research directions are summarized, especially the synergistic methods combining both big data driven approaches and traditional methods.
 
 Product design based on big data is usually performed with a large amount of real data. It provides unprecedented opportunities to leverage the collective intelligence of consumers, suppliers, designers, enterprises, etc. Moreover, advanced big data processing technology can accurately obtain user requirements, product evaluation, improve the success rate, and save development cost, time, and effort. We hope that this survey can draw further attention to big data and technologies (such as NLP, speech recognition technology, GAN, and CNN) in product design and promote its development toward intelligence and automation.

\textbf{Author Contributions: }{H.Q. and J.H. conceived the conception; H.Q. conducted literature collection and manuscript writing; J.H., S.L., C.Z. and J.W. revised and polished the manuscript. All authors have read and agreed to the published version of the manuscript.

\textbf{Funding: }This work was supported by Foundation of Guizhou University of Finance and Economics under Grant Nos. 2020XQN02 and 2020YJ020, and Science and Technology Foundation of Guizhou Province under Grant No ZK[2021]337.

\textbf{Conflicts of Interest: }The authors declare no conflict of interest.


\bibliographystyle{unsrt}  
\bibliography{references} 

\begin{thebibliography}{100}

\bibitem{RN263}
Vicky~M Story, Nathaniel Boso, and John~W Cadogan.
\newblock The form of relationship between firm‐level product innovativeness
  and new product performance in developed and emerging markets.
\newblock {\em Journal of Product Innovation Management}, 32(1):45--64, 2015.

\bibitem{RN264}
Lingguo Bu, Chun~Hsien Chen, Kam~KH Ng, Pai Zheng, Guijun Dong, and Heshan Liu.
\newblock A user-centric design approach for smart product-service systems
  using virtual reality: a case study.
\newblock {\em Journal of Cleaner Production}, 280:1--36, 2021.

\bibitem{RN167}
Xiaoling Gu, Fei Gao, Min Tan, and Pai Peng.
\newblock Fashion analysis and understanding with artificial intelligence.
\newblock {\em Information Processing \& Management}, 57:102276--102292, 2020.

\bibitem{RN265}
Daniel~E O'Leary.
\newblock Artificial intelligence and big data.
\newblock {\em IEEE intelligent systems}, 28(2):96--99, 2013.

\bibitem{RN151}
Sreedhar Radhakrishnan, Varun Bharadwaj, Varun Manjunath, and Ramamoorthy
  Srinath.
\newblock Creative intelligence--automating car design studio with generative
  adversarial networks (gan).
\newblock In {\em International Cross-Domain Conference for Machine Learning
  and Knowledge Extraction}, pages 160--175. Springer, 2018.

\bibitem{RN40}
Ching~Hung Lee, Chun~Hsien Chen, and Yu~Chi Lee.
\newblock Customer requirement-driven design method and computer-aided design
  system for supporting service innovation conceptualization handling.
\newblock {\em Advanced Engineering Informatics}, 45:1--16, 2020.

\bibitem{RN268}
Tianxiong Wang and Meiyu Zhou.
\newblock A method for product form design of integrating interactive genetic
  algorithm with the interval hesitation time and user satisfaction.
\newblock {\em International Journal of Industrial Ergonomics},
  76:102901--102917, 2020.

\bibitem{RN13}
Gülin Büyükzkan and Fethullah Ger.
\newblock Application of a new combined intuitionistic fuzzy mcdm approach
  based on axiomatic design methodology for the supplier selection problem.
\newblock {\em Applied Soft Computing}, 52:1222--1238, 2017.

\bibitem{RN12}
Xiuzhen Li, Siqi Qiu, and Henry X.~G. Ming.
\newblock An integrated module-based reasoning and axiomatic design approach
  for new product design under incomplete information environment.
\newblock {\em Computers \& Industrial Engineering}, 127:63--73, 2019.

\bibitem{RN17}
Mahmoud~Z. Mistarihi, Rasha~A. Okour, and Ahmad~A. Mumani.
\newblock An integration of a qfd model with fuzzy-anp approach for determining
  the importance weights for engineering characteristics of the proposed
  wheelchair design.
\newblock {\em Applied Soft Computing}, 90:106136--106148, 2020.

\bibitem{RN16}
Chih~Hsuan Wang.
\newblock Using the theory of inventive problem solving to brainstorm
  innovative ideas for assessing varieties of phone-cameras.
\newblock {\em Computers \& Industrial Engineering}, 85:227--234, 2015.

\bibitem{RN38}
Hajime Yamashina, Takaaki Ito, and Hiroshi Kawada.
\newblock Innovative product development process by integrating qfd and triz.
\newblock {\em International Journal of Production Research}, 40(5):1031--1050,
  2002.

\bibitem{RN140}
Ying Chen, Shixiong Xia, Jiaqi Zhao, Yong Zhou, Qiang Niu, Rui Yao, and Dongjun
  Zhu.
\newblock Appearance and shape based image synthesis by conditional variational
  generative adversarial network.
\newblock {\em Knowledge-Based Systems}, 193:105450--105477, 2020.

\bibitem{RN88}
Hui Sun, Wei Guo, Hongyu Shao, and Bo~Rong.
\newblock Dynamical mining of ever-changing user requirements: A product design
  and improvement perspective.
\newblock {\em Advanced Engineering Informatics}, 46:101174--101186, 2020.

\bibitem{RN8}
Lei Wang and Zhengchao Liu.
\newblock Data-driven product design evaluation method based on multi-stage
  artificial neural network.
\newblock {\em Applied Soft Computing}, 103(11):107117, 2021.

\bibitem{RN43}
Yung-Hung Wu and Chao~Chung Ho.
\newblock Integration of green quality function deployment and fuzzy theory: a
  case study on green mobile phone design.
\newblock {\em Journal of Cleaner production}, 108:271--280, 2015.

\bibitem{RN1}
Min Dong, Xianyi Zeng, Ludovic Koehl, and Junjie Zhang.
\newblock An interactive knowledge-based recommender system for fashion product
  design in the big data environment.
\newblock {\em Information Sciences}, 540:469--488, 2020.

\bibitem{RN41}
Tomohiko Sakao.
\newblock A qfd-centred design methodology for environmentally conscious
  product design.
\newblock {\em International journal of production research},
  45(18):4143--4162, 2007.

\bibitem{RN114}
Jian Jin, Ying Liu, Ping Ji, and CK~Kwong.
\newblock Review on recent advances in information mining from big consumer
  opinion data for product design.
\newblock {\em Journal of Computing and Information Science in Engineering},
  19(1), 2019.

\bibitem{RN6}
Zhen-Song Chen, Xiao-Lu Liu, Kwai-Sang Chin, Witold Pedrycz, Kwok-Leung Tsui,
  and Miroslaw~J Skibniewski.
\newblock Online-review analysis based large-scale group decision-making for
  determining passenger demands and evaluating passenger satisfaction: Case
  study of high-speed rail system in china.
\newblock {\em Information Fusion}, 69:22--39, 2020.

\bibitem{RN5}
Zhen-Song Chen, Xiao-Lu Liu, Rosa~M Rodríguez, Xian-Jia Wang, Kwai-Sang Chin,
  Kwok-Leung Tsui, and Luis Martínez.
\newblock Identifying and prioritizing factors affecting in-cabin passenger
  comfort on high-speed rail in china: A fuzzy-based linguistic approach.
\newblock {\em Applied Soft Computing}, 95:106558--106577, 2020.

\bibitem{RN7}
Alexandros Iosifidis, Anastasios Tefas, Ioannis Pitas, and Moncef Gabbouj.
\newblock Big media data analysis.
\newblock {\em Signal Processing: Image Communication}, 59:105--108, 2017.

\bibitem{RN10}
Yingfeng Zhang, Shan Ren, Yang Liu, Tomohiko Sakao, and Donald Huisingh.
\newblock A framework for big data driven product lifecycle management.
\newblock {\em Journal of Cleaner Production}, 159:229--240, 2017.

\bibitem{RN11}
Shan Ren, Yingfeng Zhang, Yang Liu, Tomohiko Sakao, Donald Huisingh, and
  Cecilia M. V.~B Almeida.
\newblock A comprehensive review of big data analytics throughout product
  lifecycle to support sustainable smart manufacturing: A framework, challenges
  and future research directions.
\newblock {\em Journal of Cleaner Production}, 210:1343--1365, 2019.

\bibitem{RN9}
Xianyu Zhang, Xinguo Ming, and Dao Yin.
\newblock Application of industrial big data for smart manufacturing in product
  service system based on system engineering using fuzzy dematel.
\newblock {\em Journal of Cleaner Production}, 265:121863--121888, 2020.

\bibitem{RN269}
Yingfeng Zhang, Shan Ren, Yang Liu, and Shubin Si.
\newblock A big data analytics architecture for cleaner manufacturing and
  maintenance processes of complex products.
\newblock {\em Journal of cleaner production}, 142:626--641, 2017.

\bibitem{RN39}
Jose~A Carnevalli and Paulo~Cauchick Miguel.
\newblock Review, analysis and classification of the literature on qfd—types
  of research, difficulties and benefits.
\newblock {\em International Journal of Production Economics}, 114(2):737--754,
  2008.

\bibitem{RN57}
Runliang Dou, Yubo Zhang, and Guofang Nan.
\newblock Application of combined kano model and interactive genetic algorithm
  for product customization.
\newblock {\em Journal of Intelligent Manufacturing}, 30(7):2587--2602, 2019.

\bibitem{RN2}
Yan LI; Jie WANG;Xianglong LI;Wu~ZHAO;Wei HU.
\newblock Creative thinking and computer aided product innovation.
\newblock {\em Computer Integrated Manufacturing System}, 9(12):1092--1096,
  2003.

\bibitem{RN15}
Sonal Keshwani, Torben~Anker Lena, Saeema Ahmed-Kristense, and Amaresh
  Chakrabart.
\newblock Comparing novelty of designs from biological-inspiration with those
  from brainstorming.
\newblock {\em Journal of Engineering Design}, 28:654--680, 2017.

\bibitem{RN14}
Barry~Matthew Kudrowitz and David Wallace.
\newblock Assessing the quality of ideas from prolific, early-stage product
  ideation.
\newblock {\em Journal of Engineering Design}, 24(2):120--139, 2013.

\bibitem{bonnardel2020brainstorming}
Nathalie Bonnardel and John Didier.
\newblock Brainstorming variants to favor creative design.
\newblock {\em Applied ergonomics}, 83:102987, 2020.

\bibitem{youn2015invention}
Hyejin Youn, Deborah Strumsky, Luis~MA Bettencourt, and Jos{\'e} Lobo.
\newblock Invention as a combinatorial process: evidence from us patents.
\newblock {\em Journal of the Royal Society interface}, 12(106):20150272, 2015.

\bibitem{lai2006user}
Hsin-Hsi Lai, Yang-Cheng Lin, Chung-Hsing Yeh, and Chien-Hung Wei.
\newblock User-oriented design for the optimal combination on product design.
\newblock {\em International Journal of Production Economics}, 100(2):253--267,
  2006.

\bibitem{zarraonandia2017using}
Telmo Zarraonandia, Paloma Diaz, and Ignacio Aedo.
\newblock Using combinatorial creativity to support end-user design of digital
  games.
\newblock {\em Multimedia Tools and Applications}, 76(6):9073--9098, 2017.

\bibitem{2012A}
Tomohiko Sakao and Mattias Lindahl.
\newblock A value based evaluation method for product/service system using
  design information.
\newblock {\em CIRP Annals - Manufacturing Technology}, 61(1):51--54, 2012.

\bibitem{RN18}
Joana Vieira, Joana Maria~A. Osório, Sandra Mouta, Pedro Delgado, Aníbal
  Portinha, José~Filipe Meireles, and Jorge~Almeida Santos.
\newblock Kansei engineering as a tool for the design of in-vehicle rubber
  keypads.
\newblock {\em Applied Ergonomics}, 61(1):1--11, 2017.

\bibitem{RN19}
Nagamachi and M.
\newblock Kansei engineering in consumer product design.
\newblock {\em Ergonomics in Design the Quarterly of Human Factors
  Applications}, 10(2):5--9, 2016.

\bibitem{RN20}
Mitsuo Nagamachi.
\newblock Kansei engineering: A new ergonomic consumer-oriented technology for
  product development.
\newblock {\em international journal of industrial ergonomics}, 15(1):3--11,
  1995.

\bibitem{RN22}
Mitsuo Nagamachi.
\newblock Successful points of kansei product development.
\newblock In {\em 7th International Conference on Kansei Engineering \& Emotion
  Research}, pages 177--187. Linköping University Electronic Press.

\bibitem{RN21}
Mitsuo Nagamachi and Anitawati~Mohd Lokman.
\newblock {\em Innovations of Kansei engineering}.
\newblock CRC Press, Boca Raton, US, 2016.

\bibitem{RN23}
Shigekazu Ishihara, MITSUO NAGAMACHI, SIMON SCHÜTTE, and JÖRGEN EKLUND.
\newblock {\em Affective meaning: The kansei engineering approach}, pages
  477--496.
\newblock Elsevier, 2008.

\bibitem{RN25}
Simon Schütte.
\newblock {\em Designing feelings into products: Integrating kansei engineering
  methodology in product development}.
\newblock Thesis, 2002.

\bibitem{RN26}
Simon Schütte.
\newblock {\em Engineering emotional values in product design: Kansei
  engineering in development}.
\newblock Thesis, 2005.

\bibitem{RN24}
Simon Schütte and Jörgen Eklund.
\newblock Design of rocker switches for work-vehicles—an application of
  kansei engineering.
\newblock {\em Applied ergonomics}, 36(5):557--567, 2005.

\bibitem{RN27}
Llu{\'\i}s Marco~Almagro, Xavier Tort-Martorell~Llabr{\'e}s, Simon Sch{\"u}tte,
  et~al.
\newblock A discussion on the selection of prototypes for kansei engineering
  study.
\newblock 2016.

\bibitem{RN28}
Simon~TW Schütte*, Jörgen Eklund, Jan~RC Axelsson, and Mitsuo Nagamachi.
\newblock Concepts, methods and tools in kansei engineering.
\newblock {\em Theoretical Issues in Ergonomics Science}, 5(3):214--231, 2004.

\bibitem{RN29}
Shigekazu Ishihara, Mitsuo Nagamachi, and Toshio Tsuchiya.
\newblock Development of a kansei engineering artificial intelligence
  sightseeing application.
\newblock In {\em International Conference on Applied Human Factors and
  Ergonomics}, pages 312--322. Springer, 2018.

\bibitem{RN30}
Taufik Djatna and Wenny~Dwi Kurniati.
\newblock A system analysis and design for packaging design of powder shaped
  fresheners based on kansei engineering.
\newblock {\em Procedia Manufacturing}, 4:115--123, 2015.

\bibitem{RN31}
Meng~Dar Shieh and Yu~En Yeh.
\newblock Developing a design support system for the exterior form of running
  shoes using partial least squares and neural networks.
\newblock {\em Computers \& Industrial Engineering}, 65(4):704--718, 2013.

\bibitem{RN33}
Qing~Xing Qu and Fu~Guo.
\newblock Can eye movements be effectively measured to assess product design?:
  Gender differences should be considered.
\newblock {\em International Journal of Industrial Ergonomics}, 72:281--289,
  2019.

\bibitem{RN32}
Fuqian Shi, Nilanjan Dey, Amira~S. Ashour, Dimitra Sifaki-Pistolla, and
  R.~Simon Sherratt.
\newblock Meta-kansei modeling with valence-arousal fmri dataset of brain.
\newblock {\em Cognitive Computation}, 11:227--240, 2019.

\bibitem{RN34}
Kemal~Mert Dogan, Hiromasa Suzuki, and Erkan Gunpinar.
\newblock Eye tracking for screening design parameters in adjective-based
  design of yacht hull.
\newblock {\em Ocean Engineering}, 166:262--277, 2018.

\bibitem{RN35}
Wangqun Xiao and Jianxin Cheng.
\newblock Perceptual design method for smart industrial robots based on virtual
  reality and synchronous quantitative physiological signals.
\newblock {\em International Journal of Distributed Sensor Networks},
  16(5):1--15, 2020.

\bibitem{RN36}
Tao Hu, Qingsheng Xie, Qingni Yuan, Jian Lv, and Qiaoqiao Xiong.
\newblock Design of ethnic patterns based on shape grammar and artificial
  neural network.
\newblock {\em Alexandria Engineering Journal}, 60(1):1601--1625, 2021.

\bibitem{RN55}
Noriaki KANO, Nobuhiko SERAKU, Fumio TAKAHASHI, and Shin-ichi TSUJI.
\newblock Attractive quality and must-be quality.
\newblock {\em Journal of The Japanese Society for Quality Control},
  14(2):147--156, 1984.

\bibitem{RN56}
Shwetank Avikal, Rajeev Jain, and PK~Mishra.
\newblock A kano model, ahp and m-topsis method-based technique for disassembly
  line balancing under fuzzy environment.
\newblock {\em Applied Soft Computing}, 25:519--529, 2014.

\bibitem{RN58}
Mei-Ling Yao, Ming-Chuen Chuang, and Chun-Cheng Hsu.
\newblock The kano model analysis of features for mobile security applications.
\newblock {\em Computers \& Security}, 78:336--346, 2018.

\bibitem{RN59}
Shwetank Avikal, Rohit Singh, and Rashmi Rashmi.
\newblock Qfd and fuzzy kano model based approach for classification of
  aesthetic attributes of suv car profile.
\newblock {\em Journal of Intelligent Manufacturing}, 31(2):271--284, 2020.

\bibitem{RN60}
Maria~Grazia Violante and Enrico Vezzetti.
\newblock Kano qualitative vs quantitative approaches: An assessment framework
  for products attributes analysis.
\newblock {\em Computers in Industry}, 86:15--25, 2017.

\bibitem{RN61}
Lina He, Wenyan Song, Zhenyong Wu, Zhitao Xu, Maokuan Zheng, and Xinguo Ming.
\newblock Quantification and integration of an improved kano model into qfd
  based on multi-population adaptive genetic algorithm.
\newblock {\em Computers \& Industrial Engineering}, 114:183--194, 2017.

\bibitem{RN64}
Xiuli Geng and Xuening Chu.
\newblock A new importance–performance analysis approach for customer
  satisfaction evaluation supporting pss design.
\newblock {\em Expert Systems with Applications}, 39(1):1492--1502, 2012.

\bibitem{RN62}
Yu-Cheng Lee, Liang-Chyau Sheu, and Yuan-Gan Tsou.
\newblock Quality function deployment implementation based on fuzzy kano model:
  An application in plm system.
\newblock {\em Computers \& Industrial Engineering}, 55(1):48--63, 2008.

\bibitem{RN65}
Mazaher Ghorbani, S~Mohammad~Arabzad, and Arash Shahin.
\newblock A novel approach for supplier selection based on the kano model and
  fuzzy mcdm.
\newblock {\em International Journal of Production Research},
  51(18):5469--5484, 2013.

\bibitem{RN67}
Chun-Chih Chen and Ming-Chuen Chuang.
\newblock Integrating the kano model into a robust design approach to enhance
  customer satisfaction with product design.
\newblock {\em International journal of production economics}, 114(2):667--681,
  2008.

\bibitem{RN68}
Cigdem Basfirinci and Amitava Mitra.
\newblock A cross cultural investigation of airlines service quality through
  integration of servqual and the kano model.
\newblock {\em Journal of Air Transport Management}, 42:239--248, 2015.

\bibitem{RN262}
Jiayin Qi, Zhenping Zhang, Seongmin Jeon, and Yanquan Zhou.
\newblock Mining customer requirements from online reviews: A product
  improvement perspective.
\newblock {\em Information \& Management}, 53(8):951--963, 2016.

\bibitem{RN69}
Valerio Bellandi, Paolo Ceravolo, and Maryam Ehsanpour.
\newblock A case study in smart healthcare platform design.
\newblock In {\em IEEE World Congress on Services}, pages 7--12. IEEE.

\bibitem{RN37}
B~Almannai, R~Greenough, and J~Kay.
\newblock A decision support tool based on qfd and fmea for the selection of
  manufacturing automation technologies.
\newblock {\em Robotics and Computer-Integrated Manufacturing}, 24(4):501--507,
  2008.

\bibitem{RN42}
Hong~Bin Yan, Xiang~Sheng Meng, Tieju Ma, and Van~Nam Huynh.
\newblock An uncertain target-oriented qfd approach to service design based on
  service standardization with an application to bank window service.
\newblock {\em IISE Transactions}, 51(11):1167--1189, 2019.

\bibitem{RN44}
Kwang-Jae Kim, Herbert Moskowitz, Anoop Dhingra, and Gerald Evans.
\newblock Fuzzy multicriteria models for quality function deployment.
\newblock {\em European Journal of Operational Research}, 121(3):504--518,
  2000.

\bibitem{RN45}
Cengiz Kahraman, Tijen Ertay, and Gülçin Büyüközkan.
\newblock A fuzzy optimization model for qfd planning process using analytic
  network approach.
\newblock {\em European journal of operational research}, 171(2):390--411,
  2006.

\bibitem{RN46}
Yu-Hui Wang, Ching-Hung Lee, and Amy~JC Trappey.
\newblock Service design blueprint approach incorporating triz and service qfd
  for a meal ordering system: A case study.
\newblock {\em Computers \& Industrial Engineering}, 107:388--400, 2017.

\bibitem{RN49}
Mehtap Dursun and E~Ertugrul Karsak.
\newblock A qfd-based fuzzy mcdm approach for supplier selection.
\newblock {\em Applied Mathematical Modelling}, 37(8):5864--5875, 2013.

\bibitem{RN47}
Ming Li, Lei Jin, and Jun Wang.
\newblock A new mcdm method combining qfd with topsis for knowledge management
  system selection from the user's perspective in intuitionistic fuzzy
  environment.
\newblock {\em Applied soft computing}, 21:28--37, 2014.

\bibitem{RN48}
Hao-Tien Liu.
\newblock Product design and selection using fuzzy qfd and fuzzy mcdm
  approaches.
\newblock {\em Applied Mathematical Modelling}, 35(1):482--496, 2011.

\bibitem{RN50}
Morteza Yazdani, Prasenjit Chatterjee, Edmundas~Kazimieras Zavadskas, and
  Sarfaraz~Hashemkhani Zolfani.
\newblock Integrated qfd-mcdm framework for green supplier selection.
\newblock {\em Journal of Cleaner Production}, 142:3728--3740, 2017.

\bibitem{RN51}
Xu~Wang, Hong Fang, and Wenyan Song.
\newblock Technical attribute prioritisation in qfd based on cloud model and
  grey relational analysis.
\newblock {\em International Journal of Production Research},
  58(19):5751--5768, 2020.

\bibitem{RN52}
Morteza Yazdani, Cengiz Kahraman, Pascale Zarate, and Sezi~Cevik Onar.
\newblock A fuzzy multi attribute decision framework with integration of qfd
  and grey relational analysis.
\newblock {\em Expert Systems with Applications}, 115:474--485, 2019.

\bibitem{RN53}
Lian-Yin Zhai, Li-Pheng Khoo, and Zhao-Wei Zhong.
\newblock A rough set based qfd approach to the management of imprecise design
  information in product development.
\newblock {\em Advanced Engineering Informatics}, 23(2):222--228, 2009.

\bibitem{RN54}
Lian-Yin Zhai, Li~Pheng Khoo, and Zhao-Wei Zhong.
\newblock Towards a qfd-based expert system: A novel extension to fuzzy qfd
  methodology using rough set theory.
\newblock {\em Expert Systems with Applications}, 37(12):8888--8896, 2010.

\bibitem{moussa2017reviewing}
Fatima Zahra~Ben Moussa, Ivana Rasovska, S{\'e}bastien Dubois, Roland De~Guio,
  and Rachid Benmoussa.
\newblock Reviewing the use of the theory of inventive problem solving (triz)
  in green supply chain problems.
\newblock {\em Journal of cleaner production}, 142:2677--2692, 2017.

\bibitem{ai2020low}
Xianfeng Ai, Zhigang Jiang, Hua Zhang, and Yan Wang.
\newblock Low-carbon product conceptual design from the perspectives of
  technical system and human use.
\newblock {\em Journal of Cleaner Production}, 244:118819, 2020.

\bibitem{RN71}
CL~Philip Chen and Chun-Yang Zhang.
\newblock Data-intensive applications, challenges, techniques and technologies:
  A survey on big data.
\newblock {\em Information sciences}, 275:314--347, 2014.

\bibitem{RN72}
Nusrat~J Shoumy, Li-Minn Ang, Kah~Phooi Seng, DM~Motiur Rahaman, and Tanveer
  Zia.
\newblock Multimodal big data affective analytics: A comprehensive survey using
  text, audio, visual and physiological signals.
\newblock {\em Journal of Network and Computer Applications},
  149:102447--1024482, 2020.

\bibitem{toffler1980third}
Alvin Toffler.
\newblock The third wave/alvin toffler.
\newblock {\em New York: Morrow}, 544, 1980.

\bibitem{RN73}
John Gantz and David Reinsel.
\newblock Extracting value from chaos.
\newblock {\em IDC iview}, 1142(2011):1--12, 2011.

\bibitem{RN76}
Anass Fahfouh, Jamal Riffi, Mohamed~Adnane Mahraz, Ali Yahyaouy, and Hamid
  Tairi.
\newblock Pv-dae: A hybrid model for deceptive opinion spam based on neural
  network architectures.
\newblock {\em Expert Systems with Applications}, 157:113517, 2020.

\bibitem{RN74}
Zhuo Wang and Qian Chen.
\newblock Monitoring online reviews for reputation fraud campaigns.
\newblock {\em Knowledge-Based Systems}, 195:1--12, 2020.

\bibitem{RN75}
Lan You, Qingxi Peng, Zenggang Xiong, Du~He, Meikang Qiu, and Xuemin Zhang.
\newblock Integrating aspect analysis and local outlier factor for intelligent
  review spam detection.
\newblock {\em Future Generation Computer Systems}, 102:163--172, 2020.

\bibitem{RN77}
Z~Li, ZG~Tian, JW~Wang, WM~Wang, and GQ~Huang.
\newblock Dynamic mapping of design elements and affective responses: a machine
  learning based method for affective design.
\newblock {\em Journal of Engineering Design}, 29(7):358--380, 2018.

\bibitem{RN78}
Jianxin~Roger Jiao, Yiyang Zhang, and Martin Helander.
\newblock A kansei mining system for affective design.
\newblock {\em Expert Systems with Applications}, 30(4):658--673, 2006.

\bibitem{RN79}
Amir Gandomi and Murtaza Haider.
\newblock Beyond the hype: Big data concepts, methods, and analytics.
\newblock {\em International journal of information management},
  35(2):137--144, 2015.

\bibitem{RN129}
Joao~P Carvalho, Hugo Rosa, Gaspar Brogueira, and Fernando Batista.
\newblock Misnis: An intelligent platform for twitter topic mining.
\newblock {\em Expert Systems with Applications}, 89:374--388, 2017.

\bibitem{RN128}
Raymond~YK Lau, Chunping Li, and Stephen~SY Liao.
\newblock Social analytics: Learning fuzzy product ontologies for
  aspect-oriented sentiment analysis.
\newblock {\em Decision Support Systems}, 65:80--94, 2014.

\bibitem{RN130}
Yao Liu, Cuiqing Jiang, and Huimin Zhao.
\newblock Using contextual features and multi-view ensemble learning in product
  defect identification from online discussion forums.
\newblock {\em Decision Support Systems}, 105:1--12, 2018.

\bibitem{RN127}
Yongtae Park and Sungjoo Lee.
\newblock How to design and utilize online customer center to support new
  product concept generation.
\newblock {\em Expert Systems with Applications}, 38(8):10638--10647, 2011.

\bibitem{RN131}
Long Ren, Bin Zhu, and Zeshui Xu.
\newblock Data-driven fuzzy preference analysis from an optimization
  perspective.
\newblock {\em Fuzzy Sets and Systems}, 377:85--101, 2019.

\bibitem{RN123}
Hong Hong, Di~Xu, G~Alan Wang, and Weiguo Fan.
\newblock Understanding the determinants of online review helpfulness: A
  meta-analytic investigation.
\newblock {\em Decision Support Systems}, 102:1--11, 2017.

\bibitem{RN95}
Hye-Jin Min and Jong~C Park.
\newblock Identifying helpful reviews based on customer’s mentions about
  experiences.
\newblock {\em Expert Systems with Applications}, 39(15):11830--11838, 2012.

\bibitem{RN132}
Jaewoong Choi, Janghyeok Yoon, Jaemin Chung, Byoung-Youl Coh, and Jae-Min Lee.
\newblock Social media analytics and business intelligence research: A
  systematic review.
\newblock {\em Information Processing \& Management}, 57(6):102279--102298,
  2020.

\bibitem{RN85}
Shengsheng Xiao, Chih-Ping Wei, and Ming Dong.
\newblock Crowd intelligence: Analyzing online product reviews for preference
  measurement.
\newblock {\em Information \& Management}, 53(2):169--182, 2016.

\bibitem{RN118}
Kexin Zhao, Antonis~C Stylianou, and Yiming Zheng.
\newblock Sources and impacts of social influence from online anonymous user
  reviews.
\newblock {\em Information \& Management}, 55(1):16--30, 2018.

\bibitem{RN105}
Anthony~JT Lee, Fu-Chen Yang, Chao-Hung Chen, Chun-Sheng Wang, and Chih-Yuan
  Sun.
\newblock Mining perceptual maps from consumer reviews.
\newblock {\em Decision Support Systems}, 82:12--25, 2016.

\bibitem{RN109}
Jian-Wu Bi, Yang Liu, Zhi-Ping Fan, and Erik Cambria.
\newblock Modelling customer satisfaction from online reviews using ensemble
  neural network and effect-based kano model.
\newblock {\em International Journal of Production Research},
  57(22):7068--7088, 2019.

\bibitem{RN104}
Minqing Hu and Bing Liu.
\newblock Mining opinion features in customer reviews.
\newblock In {\em AAAI}, volume~4, pages 755--760.

\bibitem{RN133}
Daekook Kang and Yongtae Park.
\newblock Review-based measurement of customer satisfaction in mobile service:
  Sentiment analysis and vikor approach.
\newblock {\em Expert Systems with Applications}, 41(4):1041--1050, 2014.

\bibitem{RN103}
Akshay Kangale, S~Krishna Kumar, Mohd~Arshad Naeem, Mark Williams, and
  Manoj~Kumar Tiwari.
\newblock Mining consumer reviews to generate ratings of different product
  attributes while producing feature-based review-summary.
\newblock {\em International Journal of Systems Science}, 47(13):3272--3286,
  2016.

\bibitem{RN96}
Yuren Wang, Xin Lu, and Yuejin Tan.
\newblock Impact of product attributes on customer satisfaction: An analysis of
  online reviews for washing machines.
\newblock {\em Electronic Commerce Research and Applications}, 29:1--11, 2018.

\bibitem{RN108}
Celestine Aguwa, Mohammad~Hessam Olya, and Leslie Monplaisir.
\newblock Modeling of fuzzy-based voice of customer for business decision
  analytics.
\newblock {\em Knowledge-Based Systems}, 125:136--145, 2017.

\bibitem{RN93}
Jiaming Zhan, Han~Tong Loh, and Ying Liu.
\newblock Gather customer concerns from online product reviews–a text
  summarization approach.
\newblock {\em Expert Systems with Applications}, 36(2):2107--2115, 2009.

\bibitem{RN86}
Nikolay Archak, Anindya Ghose, and Panagiotis~G Ipeirotis.
\newblock Deriving the pricing power of product features by mining consumer
  reviews.
\newblock {\em Management science}, 57(8):1485--1509, 2011.

\bibitem{RN82}
Darren Law, Richard Gruss, and Alan~S Abrahams.
\newblock Automated defect discovery for dishwasher appliances from online
  consumer reviews.
\newblock {\em Expert Systems with Applications}, 67:84--94, 2017.

\bibitem{RN119}
Matt Winkler, Alan~S Abrahams, Richard Gruss, and Johnathan~P Ehsani.
\newblock Toy safety surveillance from online reviews.
\newblock {\em Decision support systems}, 90:23--32, 2016.

\bibitem{RN125}
Wenhao Zhang, Hua Xu, and Wei Wan.
\newblock Weakness finder: Find product weakness from chinese reviews by using
  aspects based sentiment analysis.
\newblock {\em Expert Systems with Applications}, 39(11):10283--10291, 2012.

\bibitem{RN94}
Jian Jin, Ping Ji, and Rui Gu.
\newblock Identifying comparative customer requirements from product online
  reviews for competitor analysis.
\newblock {\em Engineering Applications of Artificial Intelligence}, 49:61--73,
  2016.

\bibitem{RN89}
Swagato Chatterjee.
\newblock Explaining customer ratings and recommendations by combining
  qualitative and quantitative user generated contents.
\newblock {\em Decision Support Systems}, 119:14--22, 2019.

\bibitem{RN111}
Zhi-Ping Fan, Guang-Ming Li, and Yang Liu.
\newblock Processes and methods of information fusion for ranking products
  based on online reviews: An overview.
\newblock {\em Information Fusion}, 60:87--97, 2020.

\bibitem{RN110}
Peide Liu and Fei Teng.
\newblock Probabilistic linguistic todim method for selecting products through
  online product reviews.
\newblock {\em Information Sciences}, 485:441--455, 2019.

\bibitem{RN113}
Yang Liu, Jian-Wu Bi, and Zhi-Ping Fan.
\newblock Ranking products through online reviews: A method based on sentiment
  analysis technique and intuitionistic fuzzy set theory.
\newblock {\em Information Fusion}, 36:149--161, 2017.

\bibitem{RN87}
Michael Siering, Amit~V Deokar, and Christian Janze.
\newblock Disentangling consumer recommendations: Explaining and predicting
  airline recommendations based on online reviews.
\newblock {\em Decision Support Systems}, 107:52--63, 2018.

\bibitem{RN84}
Jinming Zhang, Dexin Chen, and Min Lu.
\newblock Combining sentiment analysis with a fuzzy kano model for product
  aspect preference recommendation.
\newblock {\em IEEE Access}, 6:59163--59172, 2018.

\bibitem{RN106}
Min Zhang, Brandon Fan, Ning Zhang, Wenjun Wang, and Weiguo Fan.
\newblock Mining product innovation ideas from online reviews.
\newblock {\em Information Processing \& Management}, 58:102389--102402, 2021.

\bibitem{RN126}
Jian Jin, Ping Ji, and Chun~Kit Kwong.
\newblock What makes consumers unsatisfied with your products: Review analysis
  at a fine-grained level.
\newblock {\em Engineering Applications of Artificial Intelligence}, 47:38--48,
  2016.

\bibitem{RN122}
Jian Jin, Ying Liu, Ping Ji, and Hongguang Liu.
\newblock Understanding big consumer opinion data for market-driven product
  design.
\newblock {\em International Journal of Production Research},
  54(10):3019--3041, 2016.

\bibitem{RN120}
Wai~Ming Wang, JW~Wang, Zhi Li, ZG~Tian, and Eric Tsui.
\newblock Multiple affective attribute classification of online customer
  product reviews: A heuristic deep learning method for supporting kansei
  engineering.
\newblock {\em Engineering Applications of Artificial Intelligence}, 85:33--45,
  2019.

\bibitem{RN92}
Sudhanshu Kumar, Mahendra Yadava, and Partha~Pratim Roy.
\newblock Fusion of eeg response and sentiment analysis of products review to
  predict customer satisfaction.
\newblock {\em information fusion}, 52:41--52, 2019.

\bibitem{RN112}
Simon Li, Kamrun Nahar, and Benjamin~CM Fung.
\newblock Product customization of tablet computers based on the information of
  online reviews by customers.
\newblock {\em Journal of Intelligent Manufacturing}, 26(1):97--110, 2015.

\bibitem{RN99}
Qi~Li, Pengfei Li, Kezhi Mao, and Edmond Yat-Man Lo.
\newblock Improving convolutional neural network for text classification by
  recursive data pruning.
\newblock {\em Neurocomputing}, 414:143--152, 2020.

\bibitem{RN100}
Haiqing Zhang, Aicha Sekhari, Yacine Ouzrout, and Abdelaziz Bouras.
\newblock Jointly identifying opinion mining elements and fuzzy measurement of
  opinion intensity to analyze product features.
\newblock {\em Engineering Applications of Artificial Intelligence},
  47:122--139, 2016.

\bibitem{RN90}
Mohammad Tubishat, Norisma Idris, and Mohammad Abushariah.
\newblock Explicit aspects extraction in sentiment analysis using optimal rules
  combination.
\newblock {\em Future Generation Computer Systems}, 114:448--480, 2021.

\bibitem{RN124}
Changqin Quan and Fuji Ren.
\newblock Unsupervised product feature extraction for feature-oriented opinion
  determination.
\newblock {\em Information Sciences}, 272:16--28, 2014.

\bibitem{RN134}
Petar Ristoski, Petar Petrovski, Peter Mika, and Heiko Paulheim.
\newblock A machine learning approach for product matching and categorization.
\newblock {\em Semantic web}, 9(5):707--728, 2018.

\bibitem{RN81}
Zhao Fang, Qiang Zhang, Xiaoan Tang, Anning Wang, and Claude Baron.
\newblock An implicit opinion analysis model based on feature-based implicit
  opinion patterns.
\newblock {\em Artificial Intelligence Review}, 53(6):4547--4574, 2020.

\bibitem{RN83}
Duangmanee Putthividhya and Junling Hu.
\newblock Bootstrapped named entity recognition for product attribute
  extraction.
\newblock In {\em Conference on Empirical Methods in Natural Language
  Processing}, pages 1557--1567.

\bibitem{RN97}
Mohammad Tubishat, Norisma Idris, and Mohammad~AM Abushariah.
\newblock Implicit aspect extraction in sentiment analysis: Review, taxonomy,
  oppportunities, and open challenges.
\newblock {\em Information Processing \& Management}, 54(4):545--563, 2018.

\bibitem{RN98}
Hua Xu, Fan Zhang, and Wei Wang.
\newblock Implicit feature identification in chinese reviews using explicit
  topic mining model.
\newblock {\em Knowledge-Based Systems}, 76:166--175, 2015.

\bibitem{RN117}
Yin Kang and Lina Zhou.
\newblock Rube: Rule-based methods for extracting product features from online
  consumer reviews.
\newblock {\em Information \& Management}, 54(2):166--176, 2017.

\bibitem{RN102}
Minqing Hu and Bing Liu.
\newblock Mining and summarizing customer reviews.
\newblock In {\em the tenth international conference on Knowledge discovery and
  data mining}, pages 168--177. ACM.

\bibitem{RN261}
LI~Shaobo, QUAN Huafeng, HU~Jianjun, WU~Yongming, and ZHANG ansi.
\newblock Perceptual evaluation method of products based on online reviews data
  driven.
\newblock {\em Computer integrated Manufacturing Systems}, 24(3):752--762,
  2018.

\bibitem{RN91}
Wai~Ming Wang, Zhi Li, ZG~Tian, JW~Wang, and MN~Cheng.
\newblock Extracting and summarizing affective features and responses from
  online product descriptions and reviews: A kansei text mining approach.
\newblock {\em Engineering Applications of Artificial Intelligence},
  73:149--162, 2018.

\bibitem{RN101}
Yue Wang, Daniel~Y Mo, and Mitchell~M Tseng.
\newblock Mapping customer needs to design parameters in the front end of
  product design by applying deep learning.
\newblock {\em CIRP Annals}, 67(1):145--148, 2018.

\bibitem{RN249}
Li~Chen, Luole Qi, and Feng Wang.
\newblock Comparison of feature-level learning methods for mining online
  consumer reviews.
\newblock {\em Expert Systems with Applications}, 39(10):9588--9601, 2012.

\bibitem{RN248}
Rodrigo Moraes, João~Francisco Valiati, and Wilson P~GaviãO Neto.
\newblock Document-level sentiment classification: An empirical comparison
  between svm and ann.
\newblock {\em Expert Systems with Applications}, 40(2):621--633, 2013.

\bibitem{RN250}
Hai~Ha Do, PWC Prasad, Angelika Maag, and Abeer Alsadoon.
\newblock Deep learning for aspect-based sentiment analysis: a comparative
  review.
\newblock {\em Expert Systems with Applications}, 118:272--299, 2019.

\bibitem{RN251}
Yang Liu, Jian-Wu Bi, and Zhi-Ping Fan.
\newblock Multi-class sentiment classification: The experimental comparisons of
  feature selection and machine learning algorithms.
\newblock {\em Expert Systems with Applications}, 80:323--339, 2017.

\bibitem{RN253}
Liu, Yang, Jian-Wu, Fan, and Zhi-Ping.
\newblock A method for multi-class sentiment classification based on an
  improved one-vs-one (ovo) strategy and the support vector machine (svm)
  algorithm.
\newblock {\em Information Sciences}, 394:38--52, 2017.

\bibitem{RN252}
Zengcai, Hua, Yunfeng, Zhang, and Dongwen.
\newblock Chinese comments sentiment classification based on word2vec and
  svmperf.
\newblock {\em Expert Systems with Application}, 42(4):1857--1863, 2015.

\bibitem{RN254}
Iman Dehdarbehbahani, Azadeh Shakery, and Heshaam Faili.
\newblock Semi-supervised word polarity identification in resource-lean
  languages.
\newblock {\em Neural Networks}, 58:50--59, 2014.

\bibitem{RN255}
Heeryon Cho, Songkuk Kim, Jongseo Lee, and Jong-Seok Lee.
\newblock Data-driven integration of multiple sentiment dictionaries for
  lexicon-based sentiment classification of product reviews.
\newblock {\em Knowledge-Based Systems}, 71:61--71, 2014.

\bibitem{RN256}
Oscar Araque, Ganggao Zhu, and Carlos~A Iglesias.
\newblock A semantic similarity-based perspective of affect lexicons for
  sentiment analysis.
\newblock {\em Knowledge-Based Systems}, 165:346--359, 2019.

\bibitem{RN258}
Felipe Bravo-Marquez, Marcelo Mendoza, and Barbara Poblete.
\newblock Meta-level sentiment models for big social data analysis.
\newblock {\em Knowledge-based systems}, 69:86--99, 2014.

\bibitem{RN259}
Ali Yadollahi, Ameneh~Gholipour Shahraki, and Osmar~R Zaiane.
\newblock Current state of text sentiment analysis from opinion to emotion
  mining.
\newblock {\em ACM Computing Surveys}, 50(2):1--33, 2017.

\bibitem{RN260}
Yan Dang, Yulei Zhang, and Hsinchun Chen.
\newblock A lexicon-enhanced method for sentiment classification: An experiment
  on online product reviews.
\newblock {\em IEEE Intelligent Systems}, 25(4):46--53, 2009.

\bibitem{RN203}
Zhineng Chen, Shanshan Ai, and Caiyan Jia.
\newblock Structure-aware deep learning for product image classification.
\newblock {\em ACM Transactions on Multimedia Computing, Communications, and
  Applications (TOMM)}, 15(1s):1--20, 2019.

\bibitem{RN267}
Qing Li, Xiaojiang Peng, Liangliang Cao, Wenbin Du, Hao Xing, Yu~Qiao, and
  Qiang Peng.
\newblock Product image recognition with guidance learning and noisy
  supervision.
\newblock {\em Computer Vision and Image Understanding}, 196:102963--102971,
  2020.

\bibitem{RN213}
Jiwen Lu, Junlin Hu, and Jie Zhou.
\newblock Deep metric learning for visual understanding: An overview of recent
  advances.
\newblock {\em IEEE Signal Processing Magazine}, 34(6):76--84, 2017.

\bibitem{RN137}
Yi~Li, Yong Dai, Li-Jun Liu, and Hao Tan.
\newblock Advanced designing assistant system for smart design based on product
  image dataset.
\newblock In {\em International Conference on Human-Computer Interaction},
  pages 18--33. Springer.

\bibitem{RN150}
Balazs Kovacs, Peter O'Donovan, Kavita Bala, and Aaron Hertzmann.
\newblock Context-aware asset search for graphic design.
\newblock {\em IEEE transactions on visualization and computer graphics},
  25(7):2419--2429, 2018.

\bibitem{RN168}
Si~Liu, Jiashi Feng, Csaba Domokos, Hui Xu, Junshi Huang, Zhenzhen Hu, and
  Shuicheng Yan.
\newblock Fashion parsing with weak color-category labels.
\newblock {\em IEEE Transactions on Multimedia}, 16(1):253--265, 2013.

\bibitem{RN178}
Farhan Ullah, Bofeng Zhang, and Rehan~Ullah Khan.
\newblock Image-based service recommendation system: A jpeg-coefficient rfs
  approach.
\newblock {\em IEEE Access}, 8:3308--3318, 2019.

\bibitem{RN186}
Sean Bell and Kavita Bala.
\newblock Learning visual similarity for product design with convolutional
  neural networks.
\newblock {\em ACM transactions on graphics (TOG)}, 34(4):1--10, 2015.

\bibitem{RN207}
Ryan Kiros, Ruslan Salakhutdinov, and Richard~S Zemel.
\newblock Unifying visual-semantic embeddings with multimodal neural language
  models.
\newblock {\em arXiv preprint arXiv:1411.2539}, pages 1--13, 2014.

\bibitem{RN164}
Xiaobin Liu, Shiliang Zhang, Tiejun Huang, and Qi~Tian.
\newblock E2bows: An end-to-end bag-of-words model via deep convolutional
  neural network for image retrieval.
\newblock {\em Neurocomputing}, 395:188--198, 2020.

\bibitem{RN191}
Antonio Rubio, LongLong Yu, Edgar Simo-Serra, and Francesc Moreno-Noguer.
\newblock Multi-modal joint embedding for fashion product retrieval.
\newblock In {\em IEEE International Conference on Image Processing}, pages
  400--404. IEEE.

\bibitem{RN159}
Ivona Tautkute, Tomasz Trzciński, Aleksander~P Skorupa, Łukasz Brocki, and
  Krzysztof Marasek.
\newblock Deepstyle: Multimodal search engine for fashion and interior design.
\newblock {\em IEEE Access}, 7:84613--84628, 2019.

\bibitem{RN165}
Elena Andreeva, Dmitry~I Ignatov, Artem Grachev, and Andrey~V Savchenko.
\newblock Extraction of visual features for recommendation of products via deep
  learning.
\newblock In {\em International Conference on Analysis of Images, Social
  Networks and Texts}, pages 201--210. Springer.

\bibitem{RN149}
Yongwei Miao, Gaoyi Li, Chen Bao, Jiajing Zhang, and Jinrong Wang.
\newblock Clothingnet: Cross-domain clothing retrieval with feature fusion and
  quadruplet loss.
\newblock {\em IEEE Access}, 8:142669--142679, 2020.

\bibitem{RN188}
Xi~Wang, Zhenfeng Sun, Wenqiang Zhang, Yu~Zhou, and Yu-Gang Jiang.
\newblock Matching user photos to online products with robust deep features.
\newblock In {\em Proceedings of the 2016 ACM on International Conference on
  Multimedia Retrieval}, pages 7--14.

\bibitem{RN158}
Huijing Zhan, Boxin Shi, Ling-Yu Duan, and Alex~C Kot.
\newblock Deepshoe: An improved multi-task view-invariant cnn for
  street-to-shop shoe retrieval.
\newblock {\em Computer Vision And Image Understanding}, 180:23--33, 2019.

\bibitem{RN202}
Huijing Zhan, Boxin Shi, and Alex~C Kot.
\newblock Street-to-shop shoe retrieval with multi-scale viewpoint invariant
  triplet network.
\newblock In {\em IEEE International Conference on Image Processing}, pages
  1102--1106. IEEE.

\bibitem{RN155}
Shuhui Jiang, Yue Wu, and Yun Fu.
\newblock Deep bidirectional cross-triplet embedding for online clothing
  shopping.
\newblock {\em ACM Transactions on Multimedia Computing, Communications, and
  Applications (TOMM)}, 14(1):1--22, 2018.

\bibitem{RN157}
Yu-Gang Jiang, Minjun Li, Xi~Wang, Wei Liu, and Xian-Sheng Hua.
\newblock Deepproduct: Mobile product search with portable deep features.
\newblock {\em ACM Transactions on Multimedia Computing, Communications, and
  Applications (TOMM)}, 14(2):1--18, 2018.

\bibitem{RN195}
Kota Yamaguchi, M~Hadi Kiapour, Luis~E Ortiz, and Tamara~L Berg.
\newblock Retrieving similar styles to parse clothing.
\newblock {\em IEEE transactions on pattern analysis and machine intelligence},
  37(5):1028--1040, 2014.

\bibitem{RN201}
Si~Liu, Zheng Song, Guangcan Liu, Changsheng Xu, Hanqing Lu, and Shuicheng Yan.
\newblock Street-to-shop: Cross-scenario clothing retrieval via parts alignment
  and auxiliary set.
\newblock In {\em Conference on Computer Vision and Pattern Recognition}, pages
  3330--3337. IEEE.

\bibitem{RN199}
Qian Yu, Feng Liu, Yi-Zhe Song, Tao Xiang, Timothy~M Hospedales, and
  Chen-Change Loy.
\newblock Sketch me that shoe.
\newblock In {\em Conference on Computer Vision and Pattern Recognition}, pages
  799--807.

\bibitem{RN211}
Farhan Ullah, Bofeng Zhang, Rehan~Ullah Khan, Irfan Ullah, Aamir Khan, and
  Ali~Mustafa Qamar.
\newblock Visual-based items recommendation using deep neural network.
\newblock In {\em International Conference on Computing, Networks and Internet
  of Things}, pages 122--126.

\bibitem{RN147}
Xiaodan Liang, Liang Lin, Wei Yang, Ping Luo, Junshi Huang, and Shuicheng Yan.
\newblock Clothes co-parsing via joint image segmentation and labeling with
  application to clothing retrieval.
\newblock {\em IEEE Transactions on Multimedia}, 18(6):1175--1186, 2016.

\bibitem{RN190}
Xiaoling Gu, Yongkang Wong, Lidan Shou, Pai Peng, Gang Chen, and Mohan~S
  Kankanhalli.
\newblock Multi-modal and multi-domain embedding learning for fashion retrieval
  and analysis.
\newblock {\em IEEE Transactions on Multimedia}, 21(6):1524--1537, 2018.

\bibitem{RN177}
Wei-Ta Chu and Yi-Ling Wu.
\newblock Image style classification based on learnt deep correlation features.
\newblock {\em IEEE Transactions on Multimedia}, 20(9):2491--2502, 2018.

\bibitem{RN209}
Omid Poursaeed, Tomáš Matera, and Serge Belongie.
\newblock Vision-based real estate price estimation.
\newblock {\em Machine Vision and Applications}, 29(4):667--676, 2018.

\bibitem{RN171}
Tse-Yu Pan, Yi-Zhu Dai, Min-Chun Hu, and Wen-Huang Cheng.
\newblock Furniture style compatibility recommendation with cross-class triplet
  loss.
\newblock {\em Multimedia Tools and Applications}, 78(3):2645--2665, 2019.

\bibitem{RN156}
Yong-Goo Shin, Yoon-Jae Yeo, Min-Cheol Sagong, Seo-Won Ji, and Sung-Jea Ko.
\newblock Deep fashion recommendation system with style feature decomposition.
\newblock In {\em International Conference on Consumer Electronics}, pages
  301--305. IEEE.

\bibitem{RN169}
Huijing Zhan, Boxin Shi, Jiawei Chen, Qian Zheng, Ling-Yu Duan, and Alex~C Kot.
\newblock Fashion recommendation on street images.
\newblock In {\em International Conference on Image Processing}, pages
  280--284. IEEE.

\bibitem{RN185}
Haijun Zhang, Wang Huang, Linlin Liu, and Tommy~WS Chow.
\newblock Learning to match clothing from textual feature-based compatible
  relationships.
\newblock {\em IEEE Transactions on Industrial Informatics}, 16(11):6750--6759,
  2019.

\bibitem{RN210}
Zhenhen Hu, Yonggang Wen, Luoqi Liu, Jianguo Jiang, Richang Hong, Meng Wang,
  and Shuicheng Yan.
\newblock Visual classification of furniture styles.
\newblock {\em ACM Transactions on Intelligent Systems and Technology},
  8(5):1--20, 2017.

\bibitem{RN183}
Divyansh Aggarwal, Elchin Valiyev, Fadime Sener, and Angela Yao.
\newblock Learning style compatibility for furniture.
\newblock In {\em German Conference on Pattern Recognition}, pages 552--566.
  Springer.

\bibitem{RN181}
Luisa~F Polania, Mauricio Flores, Matthew Nokleby, and Yiran Li.
\newblock Learning furniture compatibility with graph neural networks.
\newblock In {\em IEEE/CVF Conference on Computer Vision and Pattern
  Recognition Workshops}, pages 366--367. Seattle, US.

\bibitem{2020Progress}
James Mountstephens and Jason Teo.
\newblock Progress and challenges in generative product design: A review of
  systems.
\newblock {\em Computers}, 9(4):80, 2020.

\bibitem{RN247}
Yabo Dan, Yong Zhao, Xiang Li, Shaobo Li, Ming Hu, and Jianjun Hu.
\newblock Generative adversarial networks (gan) based efficient sampling of
  chemical composition space for inverse design of inorganic materials.
\newblock {\em npj Computational Materials}, 6(1):1--7, 2020.

\bibitem{RN212}
Wang-Cheng Kang, Chen Fang, Zhaowen Wang, and Julian McAuley.
\newblock Visually-aware fashion recommendation and design with generative
  image models.
\newblock In {\em IEEE International Conference on Data Mining}, pages
  207--216. IEEE.

\bibitem{RN145}
Yifan Liu, Zengchang Qin, Tao Wan, and Zhenbo Luo.
\newblock Auto-painter: Cartoon image generation from sketch by using
  conditional wasserstein generative adversarial networks.
\newblock {\em Neurocomputing}, 311:78--87, 2018.

\bibitem{RN198}
Kenan~E Ak, Joo~Hwee Lim, Jo~Yew Tham, and Ashraf~A Kassim.
\newblock Semantically consistent text to fashion image synthesis with an
  enhanced attentional generative adversarial network.
\newblock {\em Pattern Recognition Letters}, 135:22--29, 2020.

\bibitem{RN184}
Taeksoo Kim, Moonsu Cha, Hyunsoo Kim, Jung~Kwon Lee, and Jiwon Kim.
\newblock Learning to discover cross-domain relations with generative
  adversarial networks.
\newblock In {\em International Conference on Machine Learning}, pages
  1857--1865. PMLR.

\bibitem{RN170}
Wei-Lin Hsiao, Isay Katsman, Chao-Yuan Wu, Devi Parikh, and Kristen Grauman.
\newblock Fashion++: Minimal edits for outfit improvement.
\newblock In {\em IEEE/CVF International Conference on Computer Vision}, pages
  5047--5056.

\bibitem{RN197}
Yining Lang, Yuan He, Jianfeng Dong, Fan Yang, and Hui Xue.
\newblock Design-gan: Cross-category fashion translation driven by landmark
  attention.
\newblock In {\em ICASSP 2020-2020 IEEE International Conference on Acoustics,
  Speech and Signal Processing (ICASSP)}, pages 1968--1972. IEEE.

\bibitem{RN189}
Jinhuan Liu, Xuemeng Song, Zhumin Chen, and Jun Ma.
\newblock Mgcm: Multi-modal generative compatibility modeling for clothing
  matching.
\newblock {\em Neurocomputing}, 414:215--224, 2020.

\bibitem{RN192}
Yong Dai, Yi~Li, and Li-Jun Liu.
\newblock New product design with automatic scheme generation.
\newblock {\em Sensing and Imaging}, 20(1):1--16, 2019.

\bibitem{RN246}
LU~Qian-Wen, TAO Qing-Chuan, ZHAO Ya-Lin, and LIU Man-Xiao.
\newblock Sketch simplification using generative adversarial networks.
\newblock {\em ACTA AUTOMATICA SINICA}, 44(5):75--89, 2018.

\bibitem{RN136}
Chunlei Chai, Jing Liao, Ning Zou, and Lingyun Sun.
\newblock A one-to-many conditional generative adversarial network framework
  for multiple image-to-image translations.
\newblock {\em Multimedia Tools and Applications}, 77(17):22339--22366, 2018.

\bibitem{RN162}
Younghoon Lee and Sungzoon Cho.
\newblock Design of semantic-based colorization of graphical user interface
  through conditional generative adversarial nets.
\newblock {\em International Journal of Human–Computer Interaction},
  36(8):699--708, 2020.

\bibitem{RN206}
Hui Ren, Jia Li, and Nan Gao.
\newblock Two-stage sketch colorization with color parsing.
\newblock {\em IEEE Access}, 8:44599--44610, 2019.

\bibitem{RN193}
Nilesh Pandey and Andreas Savakis.
\newblock Poly-gan: Multi-conditioned gan for fashion synthesis.
\newblock {\em Neurocomputing}, 414:356--364, 2020.

\bibitem{RN205}
Linlin Liu, Haijun Zhang, Yuzhu Ji, and QM~Jonathan Wu.
\newblock Toward ai fashion design: An attribute-gan model for clothing match.
\newblock {\em Neurocomputing}, 341:156--167, 2019.

\bibitem{RN172}
Morteza Rahbar, Mohammadjavad Mahdavinejad, Mohammadreza Bemanian, Amir~Hossein
  Davaie~Markazi, and Ludger Hovestadt.
\newblock Generating synthetic space allocation probability layouts based on
  trained conditional-gans.
\newblock {\em Applied Artificial Intelligence}, 33(8):689--705, 2019.

\bibitem{RN174}
Xiaolong Wang and Abhinav Gupta.
\newblock Generative image modeling using style and structure adversarial
  networks.
\newblock In {\em European conference on computer vision}, pages 318--335.
  Springer.

\bibitem{RN144}
Jianan Li, Jimei Yang, Jianming Zhang, Chang Liu, Christina Wang, and Tingfa
  Xu.
\newblock Attribute-conditioned layout gan for automatic graphic design.
\newblock {\em IEEE Transactions on Visualization and Computer Graphics}, pages
  1--10, 2020.

\bibitem{RN153}
Qingrong Cheng and Xiaodong Gu.
\newblock Cross-modal feature alignment based hybrid attentional generative
  adversarial networks for text-to-image synthesis.
\newblock {\em Digital Signal Processing}, 107:102866--102884, 2020.

\bibitem{RN173}
Scott Reed, Zeynep Akata, Xinchen Yan, Lajanugen Logeswaran, Bernt Schiele, and
  Honglak Lee.
\newblock Generative adversarial text to image synthesis.
\newblock In {\em International Conference on Machine Learning}, pages
  1060--1069. PMLR.

\bibitem{RN180}
Hongchen Tan, Xiuping Liu, Meng Liu, Baocai Yin, and Xin Li.
\newblock Kt-gan: Knowledge-transfer generative adversarial network for
  text-to-image synthesis.
\newblock {\em IEEE Transactions on Image Processing}, 30:1275--1290, 2020.

\bibitem{RN143}
Tao Xu, Pengchuan Zhang, Qiuyuan Huang, Han Zhang, Zhe Gan, Xiaolei Huang, and
  Xiaodong He.
\newblock Attngan: Fine-grained text to image generation with attentional
  generative adversarial networks.
\newblock In {\em IEEE conference on computer vision and pattern recognition},
  pages 1316--1324.

\bibitem{RN187}
Satoshi Iizuka, Edgar Simo-Serra, and Hiroshi Ishikawa.
\newblock Let there be color! joint end-to-end learning of global and local
  image priors for automatic image colorization with simultaneous
  classification.
\newblock {\em ACM Transactions on Graphics}, 35(4):1--11, 2016.

\bibitem{RN166}
Yingtao Lei, Weiwei Du, and Qinghua Hu.
\newblock Face sketch-to-photo transformation with multi-scale self-attention
  gan.
\newblock {\em Neurocomputing}, 396:13--23, 2020.

\bibitem{RN146}
Haijun Zhang, Yanfang Sun, Linlin Liu, and Xiaofei Xu.
\newblock Cascadegan: A category-supervised cascading generative adversarial
  network for clothes translation from the human body to tiled images.
\newblock {\em Neurocomputing}, 382:148--161, 2020.

\bibitem{RN179}
Phillip Isola, Jun-Yan Zhu, Tinghui Zhou, and Alexei~A Efros.
\newblock Image-to-image translation with conditional adversarial networks.
\newblock In {\em IEEE conference on computer vision and pattern recognition},
  pages 1125--1134.

\bibitem{RN175}
Jun-Yan Zhu, Philipp Krähenbühl, Eli Shechtman, and Alexei~A Efros.
\newblock Generative visual manipulation on the natural image manifold.
\newblock In {\em European conference on computer vision}, pages 597--613.
  Springer.

\bibitem{RN208}
Jun-Yan Zhu, Taesung Park, Phillip Isola, and Alexei~A Efros.
\newblock Unpaired image-to-image translation using cycle-consistent
  adversarial networks.
\newblock In {\em Proceedings of the IEEE international conference on computer
  vision}, pages 2223--2232.

\bibitem{RN204}
Nikolay Jetchev and Urs Bergmann.
\newblock The conditional analogy gan: Swapping fashion articles on people
  images.
\newblock In {\em IEEE International Conference on Computer Vision Workshops},
  pages 2287--2292.

\bibitem{RN139}
Bo~Liu, Jianhou Gan, Bin Wen, Yiping LiuFu, and Wei Gao.
\newblock An automatic coloring method for ethnic costume sketches based on
  generative adversarial networks.
\newblock {\em Applied Soft Computing}, 98:106786--106797, 2021.

\bibitem{RN161}
Sangeun Oh, Yongsu Jung, Ikjin Lee, and Namwoo Kang.
\newblock Design automation by integrating generative adversarial networks and
  topology optimization.
\newblock In {\em International Design Engineering Technical Conferences and
  Computers and Information in Engineering Conference}, page V02AT03A008.
  American Society of Mechanical Engineers.

\bibitem{RN135}
Leon~A Gatys, Alexander~S Ecker, and Matthias Bethge.
\newblock A neural algorithm of artistic style.
\newblock {\em arXiv preprint arXiv:1508.06576}, 2015.

\bibitem{RN141}
Xun Huang and Serge Belongie.
\newblock Arbitrary style transfer in real-time with adaptive instance
  normalization.
\newblock In {\em IEEE International Conference on Computer Vision}. IEEE.

\bibitem{RN194}
Huafeng Quan, Shaobo Li, and Jianjun Hu.
\newblock Product innovation design based on deep learning and kansei
  engineering.
\newblock {\em Applied Sciences}, 8(12):2397--2415, 2018.

\bibitem{RN245}
Quan Huafeng.
\newblock {\em Product design based on big data}.
\newblock Thesis, 2019.

\bibitem{RN148}
Qiang Wu, Baixue Zhu, Binbin Yong, Yongqiang Wei, Xuetao Jiang, Rui Zhou, and
  Qingguo Zhou.
\newblock Clothgan: generation of fashionable dunhuang clothes using generative
  adversarial networks.
\newblock {\em Connection Science}, pages 1--18, 2020.

\bibitem{RN142}
Kwonsang Sohn, Christine~Eunyoung Sung, Gukwon Koo, and Ohbyung Kwon.
\newblock Artificial intelligence in the fashion industry: consumer responses
  to generative adversarial network (gan) technology.
\newblock {\em International Journal of Retail \& Distribution Management},
  49(1):1--20, 2020.

\bibitem{RN182}
Yubao Sun, Jiwei Chen, Qingshan Liu, and Guangcan Liu.
\newblock Learning image compressed sensing with sub-pixel convolutional
  generative adversarial network.
\newblock {\em Pattern Recognition}, 98:107051, 2020.

\bibitem{RN244}
Ce~Wang, Zhangling Chen, Kun Shang, and Huaming Wu.
\newblock Label-removed generative adversarial networks incorporating with
  k-means.
\newblock {\em Neurocomputing}, 361:126--136, 2019.

\bibitem{RN160}
Mohammad~Hashem Faezi, Shahriar Bijani, and Ardeshir Dolati.
\newblock Degan: Decentralized generative adversarial networks.
\newblock {\em Neurocomputing}, 419:335--343, 2021.

\bibitem{RN196}
Guangcong Sun, Shifei Ding, Tongfeng Sun, and Chenglong Zhang.
\newblock Sa-capsgan: Using capsule networks with embedded self-attention for
  generative adversarial network.
\newblock {\em Neurocomputing}, 423:399--406, 2021.

\bibitem{RN228}
Shanshan Yao, Yunsheng Wang, and Baoning Niu.
\newblock An efficient cascaded filtering retrieval method for big audio data.
\newblock {\em IEEE transactions on Multimedia}, 17(9):1450--1459, 2015.

\bibitem{RN243}
Zhen YANG, Minjie XU, Zhangfeng LIU, Da~QIN, and Xiaohui YAO.
\newblock Study of audio frequency big data processing architecture and key
  technology.
\newblock {\em Telecommunication Science}, 29(11):1--5, 2013.

\bibitem{RN222}
Ching-Hung Lee, Yu-Hui Wang, and Amy~JC Trappey.
\newblock Ontology-based reasoning for the intelligent handling of customer
  complaints.
\newblock {\em Computers \& Industrial Engineering}, 84:144--155, 2015.

\bibitem{RN216}
Ying Yang, Dong-Ling Xu, Jian-Bo Yang, and Yu-Wang Chen.
\newblock An evidential reasoning-based decision support system for handling
  customer complaints in mobile telecommunications.
\newblock {\em Knowledge-Based Systems}, 162:202--210, 2018.

\bibitem{RN223}
Mustafa~Can Bingol and Omur Aydogmus.
\newblock Performing predefined tasks using the human–robot interaction on
  speech recognition for an industrial robot.
\newblock {\em Engineering Applications of Artificial Intelligence},
  95:103903--103917, 2020.

\bibitem{RN221}
Tomohiro Tanaka, Ryo Masumura, and Takanobu Oba.
\newblock Neural candidate-aware language models for speech recognition.
\newblock {\em Computer Speech \& Language}, 66:101157--101170, 2021.

\bibitem{RN220}
Yesim Dokuz and Zekeriya Tufekci.
\newblock Mini-batch sample selection strategies for deep learning based speech
  recognition.
\newblock {\em Applied Acoustics}, 171:107573--107583, 2021.

\bibitem{RN219}
Manjunath Mulimani and Shashidhar~G Koolagudi.
\newblock Extraction of mapreduce-based features from spectrograms for
  audio-based surveillance.
\newblock {\em Digital Signal Processing}, 87:1--9, 2019.

\bibitem{RN225}
Moataz El~Ayadi, Mohamed~S Kamel, and Fakhri Karray.
\newblock Survey on speech emotion recognition: Features, classification
  schemes, and databases.
\newblock {\em Pattern recognition}, 44(3):572--587, 2011.

\bibitem{RN217}
Haytham~M Fayek, Margaret Lech, and Lawrence Cavedon.
\newblock Evaluating deep learning architectures for speech emotion
  recognition.
\newblock {\em Neural Networks}, 92:60--68, 2017.

\bibitem{RN218}
Dongdong Li, Yijun Zhou, Zhe Wang, and Daqi Gao.
\newblock Exploiting the potentialities of features for speech emotion
  recognition.
\newblock {\em Information Sciences}, 548:328--343, 2021.

\bibitem{RN224}
Abdul~Malik Badshah, Jamil Ahmad, Nasir Rahim, and Sung~Wook Baik.
\newblock Speech emotion recognition from spectrograms with deep convolutional
  neural network.
\newblock In {\em International conference on platform technology and service}.
  IEEE.

\bibitem{RN242}
Mitsuo Nagamachi and Anitawati~Mohd Lokman.
\newblock {\em Kansei innovation: Practical design applications for product and
  service development}.
\newblock CRC Press, 2015.

\bibitem{RN231}
M~Shamim Hossain and Ghulam Muhammad.
\newblock Emotion recognition using deep learning approach from audio–visual
  emotional big data.
\newblock {\em Information Fusion}, 49:69--78, 2019.

\bibitem{RN227}
Yingying Jiang, Wei Li, M~Shamim Hossain, Min Chen, Abdulhameed Alelaiwi, and
  Muneer Al-Hammadi.
\newblock A snapshot research and implementation of multimodal information
  fusion for data-driven emotion recognition.
\newblock {\em Information Fusion}, 53:209--221, 2020.

\bibitem{RN232}
Jianhua Zhang, Zhong Yin, Peng Chen, and Stefano Nichele.
\newblock Emotion recognition using multi-modal data and machine learning
  techniques: A tutorial and review.
\newblock {\em Information Fusion}, 59:103--126, 2020.

\bibitem{RN236}
Guangda Li, Meng Wang, Zheng Lu, Richang Hong, and Tat-Seng Chua.
\newblock In-video product annotation with web information mining.
\newblock {\em ACM Transactions on Multimedia Computing, Communications, and
  Applications}, 8(4):1--19, 2012.

\bibitem{RN229}
Haijun Zhang, Han Guo, Xinghao Wang, Yuzhu Ji, and QM~Jonathan Wu.
\newblock Clothescounter: a framework for star-oriented clothes mining from
  videos.
\newblock {\em Neurocomputing}, 377:38--48, 2020.

\bibitem{RN239}
Haijun Zhang, Yuzhu Ji, Wang Huang, and Linlin Liu.
\newblock Sitcom-star-based clothing retrieval for video advertising: a deep
  learning framework.
\newblock {\em Neural computing and applications}, 31(11):7361--7380, 2019.

\bibitem{RN241}
Zhi-Qi Cheng, Xiao Wu, Yang Liu, and Xian-Sheng Hua.
\newblock Video2shop: Exact matching clothes in videos to online shopping
  images.
\newblock In {\em IEEE Conference on Computer Vision and Pattern Recognition},
  pages 4048--4056. IEEE.

\bibitem{RN226}
Chenglei Wu, Zhihao Tan, Zhi Wang, and Shiqiang Yang.
\newblock A dataset for exploring user behaviors in vr spherical video
  streaming.
\newblock In {\em ACM on Multimedia Systems Conference}, pages 193--198.

\bibitem{RN240}
Babak Taati, Jasper Snoek, and Alex Mihailidis.
\newblock Video analysis for identifying human operation difficulties and
  faucet usability assessment.
\newblock {\em Neurocomputing}, 100:163--169, 2013.

\bibitem{RN230}
Bo-Hao Chen, Shih-Chia Huang, and Jui-Yu Yen.
\newblock Counter-propagation artificial neural network-based motion detection
  algorithm for static-camera surveillance scenarios.
\newblock {\em Neurocomputing}, 273:481--493, 2018.

\bibitem{RN234}
Si~Liu, Xiaodan Liang, Luoqi Liu, Ke~Lu, Liang Lin, Xiaochun Cao, and Shuicheng
  Yan.
\newblock Fashion parsing with video context.
\newblock {\em IEEE Transactions on Multimedia}, 17(8):1347--1358, 2015.

\bibitem{RN235}
Haoye Dong, Xiaodan Liang, Xiaohui Shen, Bowen Wu, Bing-Cheng Chen, and Jian
  Yin.
\newblock Fw-gan: Flow-navigated warping gan for video virtual try-on.
\newblock In {\em IEEE/CVF International Conference on Computer Vision}, pages
  1161--1170. IEEE.

\bibitem{RN233}
Cheng Yi, Zhenhui Jiang, and Izak Benbasat.
\newblock Enticing and engaging consumers via online product presentations: The
  effects of restricted interaction design.
\newblock {\em Journal of Management Information Systems}, 31(4):213--242,
  2015.

\bibitem{RN237}
Shan An, Si~Liu, Zhibiao Huang, Guangfu Che, Qian Bao, Zhaoqi Zhu, Yu~Chen, and
  Dennis~Z Weng.
\newblock Rotateview: A video composition system for interactive product
  display.
\newblock {\em IEEE Transactions on Multimedia}, 21(12):3095--3105, 2019.

\end{thebibliography}






\end{document}